\documentclass[aps,prl,reprint,groupedaddress,superscriptaddress]{revtex4-2}

\usepackage{placeins}

\newcounter{suppnote}
\renewcommand{\thesuppnote}{S\arabic{suppnote}}

\newcommand{\suppnote}[2]{%
  \FloatBarrier
  \refstepcounter{suppnote}%
  \phantomsection
  \label{#1}%
  \subsection*{\thesuppnote. #2}%
}

\usepackage{anysize}
\usepackage{fancyhdr}
\usepackage{graphicx}
\usepackage{array}   % >{\raggedright...} column specifiers in tables
\usepackage{amsmath,amssymb,amsfonts}
\usepackage{braket} % provides \ket, \bra
\usepackage{booktabs}
\usepackage{multirow}
\usepackage{algorithm}      % SI pseudocode float (Supplementary Note S1)
\usepackage{algpseudocode}  % \Procedure/\State/\For/\If
\usepackage{siunitx}
\usepackage{xcolor}
\usepackage[hidelinks]{hyperref}

\graphicspath{{paper_figs/}{./}{../paper_figs/}{overleaf/paper_figs/}}

\newcommand{\figpanelref}[2]{%
  \hyperref[#1]{Fig.~\ref*{#1}(#2)}%
}

\begin{document}
\setcounter{secnumdepth}{3}

\title{Realizing Unstructured PDE Forecasting \\ on Superconducting Quantum Processors}


\author{Erik Connerty}
\email[Contact Author: ]{erikc@cec.sc.edu}
\affiliation{QodeX Quantum, Chicago, IL, USA}
\affiliation{University of South Carolina, Columbia, SC, USA}

\author{Jonathan Eastridge}
\affiliation{Naval Surface Warfare Center Panama City Division,
Panama City, FL, USA}

\author{Taylor Patti}
\affiliation{NVIDIA, Santa Clara, CA, USA}

\author{Gerasimos Angelatos}
\affiliation{RTX BBN, Cambridge, MA, USA}

\author{Ethan N. Evans}
\affiliation{QodeX Quantum, Chicago, IL, USA}


\begin{abstract}
Simulating high dimensional unstructured Partial Differential Equations (PDEs) with quantum processors is a major challenge with significant potential benefit. Existing methods have been demonstrated at small qubit counts, on simplifications of realistic problems. We implement a quantum reservoir network (QRN) for forecasting high-dimensional unstructured PDEs using up to 52 qubits on IBM superconducting hardware. Our method is gradient-free and does not suffer from the challenges of variational quantum algorithms. Instead, the QRN provides a large kernel space, up to $2^{52}$ dimensions and beyond the reach of exact classical simulation, that remains useful under current hardware noise levels. Furthermore, a partial measure-and-reset mechanism enables total circuit runtime for $\sim$24 times the qubit coherence time, so qubit lifetimes do not limit the forecast horizon. Thus, our method enables forecasting realistic unstructured nonlinear PDEs on quantum hardware. Typical difficulties associated with predicting large scale PDEs, such as transpilation, exponential shot requirements, hardware noise, and short qubit lifetimes are mitigated through careful design considerations and a dynamic observable reconstruction method. We validate our approach using a $\sim$34k cell unstructured 2D cylinder flow at Reynolds number 1000 and a challenging nonlinear Gray--Scott reaction--diffusion problem, demonstrating high fidelity prediction on CUDA-Q simulation and on 26 and 52-qubit superconducting quantum processors. Specifically, we show that adding QRN features to an autoencoder linear prediction pipeline reduces prediction error by $21\%$-$43\%$. Finally, we demonstrate closed-loop autoregressive PDE forecasting on these problems. This work demonstrates, for the first time, quantum circuits with depth beyond qubit lifetimes achieving accurate unstructured mesh forecasting on highly nonlinear PDEs.
\end{abstract}

%\keywords{}

\maketitle
\pagestyle{plain}
Simulating and forecasting Partial Differential Equations (PDEs) is an ongoing computational challenge that spans many scientific and engineering disciplines, ranging from fusion reactor development~\cite{batchelor2007simulation}, weather prediction~\cite{bauer2015quiet},
chemical reaction modeling~\cite{lu2009toward,pearsongrayscott},
and aerodynamic calculations of innovative wing designs~\cite{spalart2016cfd}.
To reduce computational burden, at realistic scales these simulations leverage so-called unstructured discretization schemes, which are unstructured graphs of polyhedrons whose density varies spatially across the problem domain~\cite{skamarock2012mpas}. Despite this and many other methods to alleviate the computational burden~\cite{berger1984amr,benner2015survey,pfaff2021meshgraphnets},
modern PDE simulations of interest are prohibitively expensive, even with the use of national-scale high performance computing infrastructure~\cite{slotnick2014cfd2030}.
Among other specialty computing techniques, quantum processor-based methods have been proposed to remediate these computational bottlenecks.

In particular, quantum methods for simulating PDEs have made marked progress in the last fifteen years, and have garnered attention along with the large and growing computational capabilities afforded by advancements in quantum computing hardware and algorithms.
Quantum algorithms for PDEs have developed along three directions~\cite{succi2023quantum,tennie2025quantum}: 
first-principles solvers that encode the underlying governing equations~\cite{harrow2009hhl,jin2024schrodingerization}, quantum physics-informed neural networks (Q-PINNs) trained to satisfy them~\cite{kyriienko2021dqc,markidis2022qpinn}, and data-driven models that learn the dynamics from trajectories~\cite{jain2024quantum,chen2025qrnnencoder}.
Each has so far been held back by differing combinations of problem structure, trainability, and hardware realizability.

First-principles methods reduce the discretized PDE to a quantum linear system~\cite{harrow2009hhl,childs2021highprecision}
or to Hamiltonian simulation~\cite{jin2024schrodingerization,an2023linear}, and treat nonlinearity through Carleman~\cite{liu2021efficient}, Koopman--von Neumann~\cite{joseph2020koopman}, or lattice-Boltzmann~\cite{sanavio2024lattice}
embeddings. Their promised exponential speedups assume efficient state preparation and full-state readout, and shrink to polynomial once these costs are counted~\cite{aaronson2015fineprint,linden2022heat}.
For nonlinear flows, efficiency guarantees further require dissipation to outweigh nonlinearity~\cite{liu2021efficient,lewis2024limitations}, leading to convergence results only on flows with low Reynolds number (the ratio of inertial to viscous forces) up to $\mathrm{Re}\sim10^{2}$~\cite{jennings2026end}.
Their encodings also assume uniform grids, leaving unstructured meshes as an open problem~\cite{deiml2024quantum,cecile2026quantum}. 
The most complete flow-past-a-cylinder demonstrations with these methods reach only $\mathrm{Re}=6$--$43$ in classical emulation~\cite{jennings2026nontrivial,su2026density}. 
Near-term variational versions~\cite{bravoprieto2023vqls,lubasch2020nonlinear}
not only keep these constraints, but add trainability limitations~\cite{larocca2024} and measurement overheads of $\sim10^{8}$ shots per cost evaluation~\cite{Sarma2024nonlinearPDE}. As a result, hardware runs have been limited to linear problems, or problems whose nonlinearity is evaluated classically, on at most about a thousand grid points with fewer than twenty gate-based qubits~\cite{Ali2023Poisson,SONG2025NavierStokes,meng2024simulating,bastidazamora2026quantum}.

Q-PINNs instead represent the solution $u(x,t)$ as the measured output of a parameterized circuit trained on the PDE residual~\cite{kyriienko2021dqc,markidis2022qpinn,trahan2024qpinn}. This removes the grid and has reached three-dimensional geometries~\cite{sedykh2024hqpinn}, but despite their large Hilbert space, these models have shown mixed gains over classical PINNs~\cite{farea2025qcpinn,berger2025trainable}. They also inherit the barren plateaus and measurement costs of variational training~\cite{mcclean2018,larocca2024}, reported demonstrations remain in classical simulation~\cite{sedykh2024hqpinn,leong2025hybrid}, and their scalability to large turbulent hardware-executed problems remains to be established.

Data-driven methods learn the dynamics from simulated or measured trajectories, usually in a compressed latent space. Most are also variational, including quantum neural operators~\cite{jain2024quantum,xiao2025qdeeponet}, quantum recurrent networks on autoencoder latents~\cite{chen2025qrnnencoder,takagi2025mpqlstm}, and quantum Koopman and prior models~\cite{zhang2026koopmanhw,wang2026qiml,xue2026qcml}, and so typically share the training burden above. These models have reached turbulent and high-Reynolds-number data, including experimental bluff-body wakes and unstructured-mesh flows~\cite{takagi2025mpqlstm,li2025qpod,wang2026qiml}, but the closest demonstrations on quantum hardware use at most 10--15 qubits per circuit~\cite{wang2026qiml,xue2026qcml,zhang2026koopmanhw}.

In contrast, quantum reservoir computing (QRC) methods are an underexplored example of data-driven methods that appear to be more scalable than both first-principles methods and Q-PINNs, while maintaining the ability to operate on compressed representations of unstructured meshes. They are non-variational by design, and thus are a departure from other data-driven methods. Crucially, the quantum circuit is fixed and only a classical linear readout is trained~\cite{mujal_opportunities_2021}. The exponentially large Hilbert space of the register serves as a rich nonlinear feature map, while all learning happens classically, so there are no quantum gradients to estimate and training reduces to linear regression. QRC therefore offers access to a kernel space that grows exponentially with qubit number, without the trainability problems of variational circuits.

QRC methods have been demonstrated on several hardware platforms~\cite{chen2020temporal,kornjača2024largescalequantumreservoirlearning,krisnanda_experimental_2025}. Feedback-driven variants re-inject measured outputs~\cite{feedback_PRXQuantum.5.040325,Murauer_2025,monomi2025feedbackenhancedquantumreservoircomputing}, while recurrent variants instead use weak measurements and non-unital dynamics to enable a persistent fading memory in the internal quantum reservoir state~\cite{yasuda2023quantumreservoircomputingrepeated,Mujal2023, Hu2024}.
A recent hardware demonstration of the feedback-driven variant \cite{dao_breaking_2026} established that QRC can be effectively scaled to greater than $100$ qubit processors and take advantage of this dimensionality for improved performance; such approaches are better suited however to static tasks, as the overhead of re-encoding previous inputs or outputs each step limits the complexity of PDEs that can practically be handled.  In contrast, recurrent QRC exploiting partial measurements and resets on superconducting processors have demonstrated tunable fading memories matched to the problem of interest ~\cite{connerty2026controllablequantummemorycapacity}, and have been proven both in theory and experiment to operate entirely unconstrained by qubit $T_1/T_2$ lifetimes~\cite{Hu2024, Connerty2026}

QRC has been applied to chaotic and reduced-order fluid dynamics~\cite{pfefferPhysRevResearch.4.033176,pfeffer2023rayleighbenard,ahmed2024prediction,salatino2026forecasting,Connerty2026,connerty2026leveragingmetrologicallyusefulstates}, but only in noiseless simulation or on at most about a dozen hardware qubits, and, to our knowledge, never to PDE fields on an unstructured mesh.

In this manuscript, we build on recurrent QRC by introducing a scalable method for forecasting large unstructured PDEs. We go on to demonstrate the experimental effectiveness of our technique using up to 52 qubits on IBM superconducting hardware. In particular, we use a graph neural network (GNN) autoencoder to compress PDE fields on an unstructured mesh into a low-dimensional latent state, decoupling geometry and boundary conditions from the quantum model by construction. The latent state drives a quantum reservoir network (QRN) whose partial measure-and-reset mechanism keeps the per-step circuit depth constant while allowing for persistent memory beyond QPU $T_1/T_2$ times~\cite{Hu2024,Connerty2026,connerty2026controllablequantummemorycapacity}. Sampled observables then feed a linear readout that predicts the latent state at later time steps. Because no circuit parameter is trained, the approach removes both the grid constraint of first-principles solvers and the training burden of variational solvers and Q-PINNs. 

We further introduce a dynamic observable reconstruction method, which mitigates the exponential shot cost of reading out larger registers. It retains Pauli observables only up to a maximum weight, chosen to suit the prediction task and the available shot budget. Finally we introduce a closed-loop autoregressive prediction protocol, realizable on current hardware, that allows the network to forecast from its own predictions.
Together, these allow us to forecast a 2D cylinder flow at $\mathrm{Re}_D=1000$ on a 33{,}824-cell unstructured mesh and a nonlinear Gray--Scott reaction--diffusion system, both in CUDA-Q simulation and on 26- and 52-qubit IBM processors. 
Adding the QRN features to the same latent-space linear readout reduces one-step and observer prediction error by 21--43\%, 
with circuits running for up to ${\sim}28\times$ the median qubit $T_1$ and up to ${\sim}24\times$ the median qubit $T_2$. The central result is that this large quantum kernel space, run on noisy hardware and with no circuit training, still adds predictive value to the PDE forecast. To our knowledge, this is the first PDE forecast on an unstructured, body-fitted mesh performed on quantum hardware, and the first closed-loop autoregressive PDE rollout executed on a quantum processor.

\section{\label{sec:model}Model Description}

\begin{figure*}[!t]
    \centering
    \includegraphics[
        width=\textwidth,
        trim=0 0 0 0,
        clip
    ]{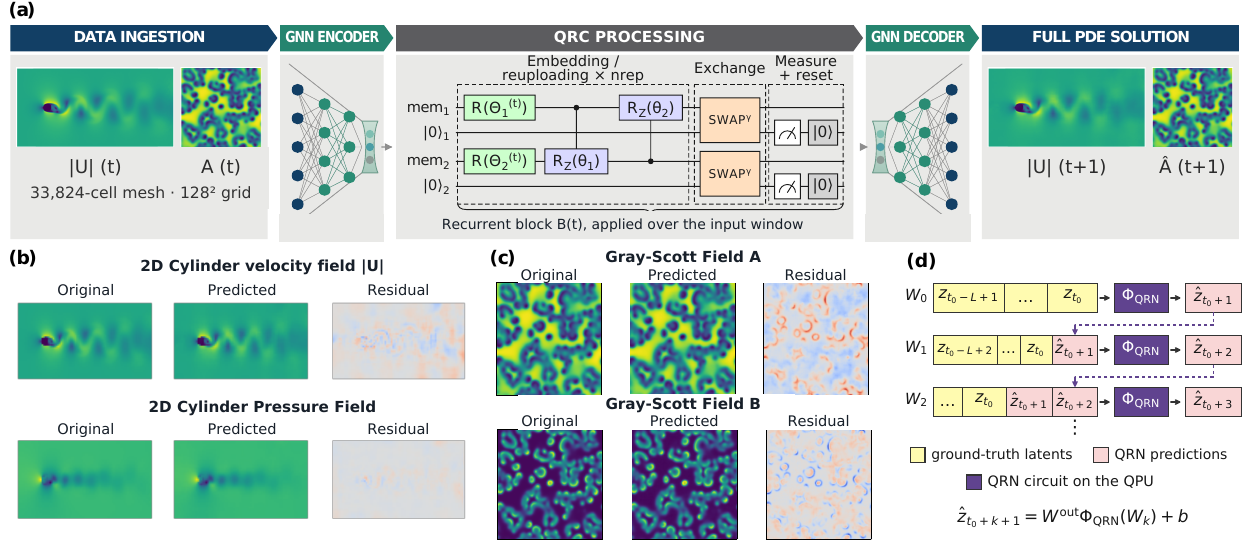}
    \caption{QRN architecture, hardware predictions, and closed-loop rollout protocol.
(a) End-to-end quantum--classical architecture. A graph autoencoder compresses PDE fields into latent states that drive the QRN, whose recurrent block $B(t)$ acts on a memory register (rows $\mathrm{mem}_i$) and a readout register (rows $|0\rangle_i$) in three stages. In the embedding stage, the latent state is re-uploaded $n_{\mathrm{rep}}$ times as rotations $R(\Theta(t))$ on the memory qubits, interleaved with controlled-$R_Z$ gates. A partial-SWAP exchange of strength $\gamma$ weakly entangles each memory qubit to its readout partner, and a measure-and-reset of the readout qubits then applies a back-action on the memory register, together forming a tunable amplitude-damping channel on the memory register. Thus, the reservoir forgets old inputs at a rate set by $\gamma$ and can operate far beyond qubit coherence times. Sampled observables feed a trained linear readout to predict the next latent state, which is decoded into PDE fields. 
(b) Reference fields, hardware predictions, and residuals for 2D cylinder flow using 26 qubits, showing velocity magnitude from one-step forecasting (top) and pressure from observer prediction (bottom). 
(c) Corresponding comparisons for Gray--Scott species $A$ (top) and $B$ (bottom) using 52 qubits on IBM QPUs. 
(d) Closed-loop rollout using a sliding latent window $W$ of fixed length $L$. Each predicted latent is appended and the oldest state is removed to form the input window for the next prediction.}
\label{fig:qrn_overview}
\end{figure*}

The end-to-end pipeline is shown in \figpanelref{fig:qrn_overview}{a}. A pretrained graph autoencoder (GAE) encodes each PDE snapshot into a latent vector 
$z_t \in \mathbb{R}^{d_z}$, which is embedded into a quantum circuit using a data reuploading block~\cite{P_rez_Salinas_2020,schuld2021effect,express_Xiong2025}, which performs a nonlinear transform on the inputs into Hilbert space~\cite{Govia_2022,express_schutte2025expressivitylimitsquantumreservoir}.
After some evolution, measurements of the readout qubits are used to obtain a vector of observables $x^{(t)}$, as detailed in Sec.~\ref{subsec:dyn_obser}. Those observables are concatenated with the current latent state as a ResNet-style ``skip'' connection~\cite{resnet_He_2016_CVPR} to form the readout features $\Phi_{\mathrm{QRN}}(z_t)=[x^{(t)};z_t]$. A regularized linear readout is then trained to predict the next latent state according to
\begin{equation}
    \hat z_{t+1}
    =
    W^{\mathrm{out}}\,\Phi_{\mathrm{QRN}}(z_t) + b,
\end{equation}
where $W^{\mathrm{out}}$ and $b$ are the learned weights and
biases, so the observables provide
features beyond the direct linear dependence on the current
latent state. Finally, the predicted latent state is passed
through the GAE decoder to reconstruct the predicted PDE field. Importantly, the GAE is trained completely independently of the QRN, and is only tasked with learning the forward and inverse mapping of unstructured PDE discretizations to latent space vectors. The QRN is tasked to forecast the latent space dynamics of each system, which we observe remain nonlinear dynamics as evidenced by the traces in Sec~\ref{si:traces} and as reflected by the error of the linear baseline (Table~\ref{tab:summary}).

Forecasting these latent dynamics requires the reservoir to have a \textit{fading memory}, which the QRN implements with a non-unital map acting on two registers, denoted \textit{memory} and \textit{readout}. In each embedding layer, the memory qubits receive Euler-angle rotations parameterized by the current latent vector, followed by fixed-angle entangling gates along the memory register; repeating this layer forms the data-reuploading block. A memory-exchange layer of partial-SWAP gates with strength $\gamma$ then couples each memory qubit to its readout partner, after which the readout register is measured and reset, exerting a purifying back-action on the memory register. Together these layers act as a controlled amplitude-damping channel on the memory register that, like the fading memory of classical echo-state networks, retains recent inputs while discarding older information at a rate set by $\gamma$~\cite{connerty2026controllablequantummemorycapacity}. These operations yield a memory register that continues computing far beyond qubit coherence times, allowing for QRNs with in principle \textit{arbitrary} gate depths. Earlier QRNs demonstrated over $100\times$ the $T_1$ and $T_2$ times~\cite{Connerty2026}, and the present circuits reach up to ${\sim}28\times T_1$ and ${\sim}24\times T_2$. Unlike feedback-based methods~\cite{Murauer_2025,feedback_PRXQuantum.5.040325,pfefferPhysRevResearch.4.033176,gonon2026feedback,monomi2025feedbackenhancedquantumreservoircomputing,feed_Zhu2025,feed_Paparelle2026-qj},
which carry memory by re-injecting measurement outcomes into the reservoir, the QRN creates this memory without collapse of the memory register.

\subsection{\label{subsec:dyn_obser}Dynamic Observable Reconstruction}

Quantum measurements are fundamentally samples from a probability distribution, and in practice one must estimate observables from a finite number of shots, which imposes scaling challenges if not carefully managed \cite{hu_tackling_2023}.
At each time step, measurements of the $n_R =
n_{\text{qubits}}/2$ readout qubits produce samples from a
distribution over $2^{n_R}$ outcomes. Retaining only the
$n_R$ single-qubit expectations is scalable, but discards the
multi-qubit correlations generated by the entangling circuit. We
therefore construct a tunable set of Pauli-$Z$ observables from the measured bitstrings, interpolating between single-qubit features and a full set of observables representing the complete distribution.

Concretely, the circuit produces $n_{\mathrm{shots}}$
measurement bitstrings $b=(b_1,\ldots,b_{n_R})\in\{0,1\}^{n_R}$ at each time-step $t$.
Let $c_b^{(t)}$ denote the number of times bitstring $b$ is observed.
A subset $S\subseteq\{1,\ldots,n_R\}$ specifies the support of a
Pauli-$Z$ string $Z_S=\bigotimes_{i=1}^{n_R}P_i$, where $P_i=Z$ if
$i\in S$ and $P_i=I$ otherwise. The corresponding expectation value
is estimated as
\begin{equation}
\widehat{\langle Z_S\rangle}^{(t)}
=
\frac{1}{n_{\mathrm{shots}}}
\sum_b c_b^{(t)}
\prod_{i\in S}(-1)^{b_i},
\end{equation}
where $|S|$ is the Pauli weight of $Z_S$.
These observables are joint moments of the measured binary variables, with the Pauli-weight specifying the moment order. Under independent symmetric readout errors with per-qubit bit-flip probability $p$, weight-$k$ moments are attenuated by the factor $(1-2p)^k$, making higher-weight correlations harder to distinguish from sampling noise \cite{readouterror_PhysRevA.105.032620}.  Retaining all strings through weight $K$ produces
%\begin{equation}
$
F=\sum_{k=1}^{K}\binom{n_R}{k}
$
%\end{equation}
features.
At $K=n_R$, this yields $F=2^{n_R}-1$ nontrivial
observables, matching the number of independent parameters in the $2^{n_R}$-outcome probability distribution. The omitted identity observable has the fixed expectation $\langle I^{\otimes n_R}\rangle=1$, corresponding to normalization.

Together, this complete hierarchy forms the Walsh--Hadamard representation of the measured distribution
\cite{odonnell2021analysisbooleanfunctions}. Consequently, reconstructing the full $2^{n_R}$-outcome distribution to total-variation error $\epsilon$, with probability at least $1-\delta$, requires
% \begin{equation}\label{eq:shotscaling}
$
n_{\mathrm{shots}}
=
\Theta\left(
\frac{2^{n_R}+\ln(1/\delta)}{\epsilon^2}
\right)
$
% \end{equation}
samples~\cite{canonne2020shortnotelearningdiscrete,
devroye2001combinatorial}. Notably, this implies an exponential shot-scaling requirement as more qubits are measured, limiting feasibility with larger circuits. However, as our method is a form of QRC, every observable is calculated from the same measurement record, allowing for post-hoc selection of $K$ dependent upon shot-budget and hardware noise. Truncation of the retained subset allows for the $2^{n_R}$ factor in $n_\mathrm{shots}$ to be reduced to $2^{K}$. Practically, the useful observable weight $K$ is only limited by finite sampling and by the sensitivity of higher-weight correlations to hardware noise~\cite{readouterror_PhysRevA.105.032620,
hu_tackling_2023}. We report an ablation over differing Pauli weights on both hardware and simulators in
Supp.\@ Note S5 \cite{supplemental}.

\section{\label{sec:results}Results}
Experiments are conducted in simulation using CUDA-Q
\cite{kim2023cuda} and on IBM superconducting
QPUs, using between 10,000 and 30,000 shots per circuit execution.
Detailed shot budgets and experimental parameters are provided in Supp.\@ Note S5 \cite{supplemental}. The datasets used are a 2D flow-past-a-circular-cylinder system at Reynolds number $\mathrm{Re}_D = 1000$, generated using OpenFOAM on a 33,824-cell unstructured mesh, and the Gray--Scott reaction--diffusion system from The Well dataset~\cite{NEURIPS2024_4f9a5acd} on a $128 \times 128$ uniform grid. The cylinder dataset contains 300 snapshots of velocity and pressure fields exhibiting periodic vortex shedding, while the Gray--Scott dataset contains 1,001 snapshots of two interacting species exhibiting spiral-pattern dynamics. Both datasets are highly nonlinear, which is a major challenge for quantum methods \cite{liu2021efficient,Jennings2026quantumalgorithms,gonzalez2025quantum}. As discussed in Sec.~\ref{sec:model}, the GAE then encodes the data into latent representations and that data is fed through the QRN circuit to generate features used for each predictive task. Details on the dataset and implementation are included in the Supp.\@ Note \cite{supplemental}.

The QRN is evaluated on three prediction tasks. In \emph{one-step} forecasting, the network receives the encoded dataset state $z_t$ at each time step and predicts $z_{t+h}$, where horizon $h=1$. The GAE decoder then converts into velocity fields $U_x, U_y$ for cylinder flow or species concentrations $A,B$ for Gray--Scott. In \emph{observer} prediction, the cylinder velocity field at time $t$ is used to predict the pressure field at $t+1$, without pressure supplied as input. In \emph{closed-loop} forecasting, the network starts from an encoded dataset state and predicts 10 successive time steps, feeding each predicted latent state back as the next input. One-step and closed-loop forecasting are evaluated on both datasets, while observer prediction is restricted to cylinder flow. Results beyond horizon $h=1$ are provided in Supp.\@ Note S10 \cite{supplemental}.

Table~\ref{tab:summary} summarizes the one-step and observer prediction results. 
Adding the QRN features improves on a linear baseline operating on the same latent representation in every case,
reducing normalized root mean square error (NRMSE) by approximately 21--43\%. The baseline uses the same ridge readout with the QRN's features removed, $\hat z_{t+1}=Wz_t+b$, so the improvement isolates the contribution of the QRN features in predicting the nonlinear latent dynamics. This baseline shows that the QRN features add predictive value, however establishing utility goes beyond the goals of this manuscript and would require an ablation against different classical architectures trained on the same latents.
This improvement persists on IBM hardware, although cylinder-flow errors increase relative to noiseless CUDA-Q simulation. For Gray--Scott, the reported NRMSE remains unchanged across simulation and hardware. This result persisted with increasing qubit count and different Pauli weight $K$, suggesting a plateau in predictive performance. Reconstruction accuracy also depends on the GAE's ability to represent the underlying fields. \figpanelref{fig:qrn_overview}{b} and \figpanelref{fig:qrn_overview}{c} show representative hardware reconstructions for the one-step and observer tasks. The reported NRMSE comes from the best performing Pauli weight $K$ in each of the experiments. Higher order Pauli weights are more susceptible to hardware noise and are also bounded in fidelity by the number of shots recovered from the circuit, therefore higher values of $K$ provided better performance in noiseless CUDA-Q simulation, and lower values performed best on hardware. More detail on the Pauli weight selection is given in Supp.\@ Note S5 \cite{supplemental}.

\begin{figure}[!ht]
  \centering
  \includegraphics[width=0.5\textwidth]{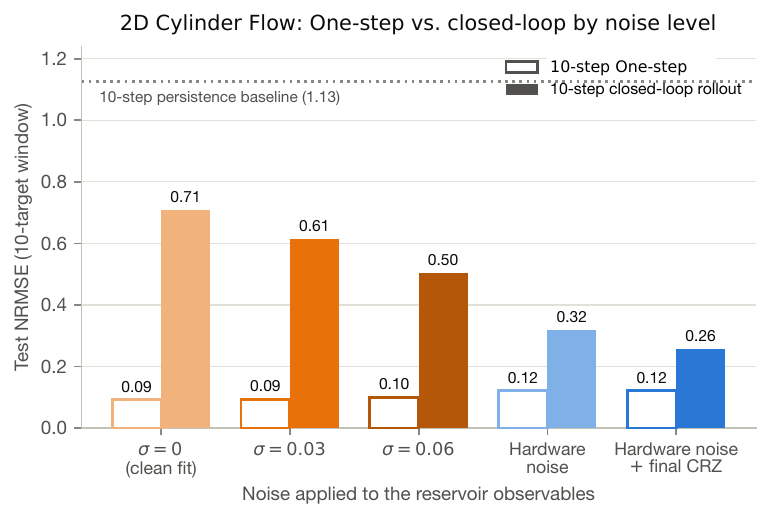}
  \caption{One-step versus closed-loop error by noise level, on
  the shared 10-step 2D-cylinder window (targets 240--249). Solid
  bars are the 10-step closed-loop rollouts. Open bars apply the same learned readout to the recorded ground-truth-driven observables over the identical targets, the 10-step one-step counterpart. The hardware and simulated runs are not seed-matched, but this discrepancy does not explain the trend seen here, as the one-step performance of the simulated runs are universally higher. A 10-step persistence model gives the baseline performance.}
  \label{fig:ar-noise-ablation}
\end{figure}

Closed-loop rollouts are performed in both simulation and hardware as a proof-of-concept, and compared to a classical baseline common to closed-loop autoregressive forecasting called \textit{persistence}~\cite{rasp2020weatherbench}. The protocol shown in \figpanelref{fig:qrn_overview}{d} re-initializes the reservoir and replays a sliding window $\mathcal{W}_k$ of $L=40$ latent states before each prediction, incorporating predicted states as the rollout proceeds. Here, $\Phi_{\mathrm{QRN}}(\mathcal{W}_k)$ denotes the readout features obtained after processing the window, including its final latent state as the skip connection. This replay washes out initial conditions and approximates the reservoir state obtained from the full input history. Replaying only the recent window avoids restarting from the beginning of the time series and requires $\mathcal{O}(LN)$ reservoir updates for an $N$-step rollout, with circuit size and shot count held fixed. Figure \ref{fig:ar-noise-ablation} compares resulting autoregressive rollouts from the 2D cylinder data against their one-step closed-loop errors. The best performing circuit was the hardware rollout run with an extra layer of CRZ gates included, which is an additional form of noise. The simulated autoregressive rollouts were universally worse than hardware, and only improved with artificial noise injection during training.  See the Supp.\@ Note S7 \cite{supplemental} for Gray--scott rollouts and details on model parameters. Surprisingly, rollout error decreases as noise increases, so rather than limiting the reservoir hardware noise appears to improve closed-loop forecasting. In this closed-loop setting, the increased noise is thought to regularize the training to make it less prone to overfitting \cite{noise_pasini2024continuousautoregressivemodelsnoise}. The authors do note that the hardware and simulated runs for this experiment did not use the same circuit seed, but this does not explain the trend seen, as the simulated circuits universally perform better at one-step prediction but worse in the autoregressive rollouts.

Finally, Table~\ref{tab:runtimes} summarizes execution times on classical GPU-accelerated simulation with CUDA-Q, and on the \texttt{ibm\_boston} quantum processor. For the 26-qubit circuits, QPU execution is about $1{,}800$--$3{,}300\times$ faster than exact CUDA-Q statevector sampling of the same circuits on a single H100 GPU. This gap is expected since each shot of a measure-and-reset circuit must be simulated as a separate trajectory. The 52-qubit circuits executed on hardware are beyond exact statevector simulation. This comparison is only between running the QRN on a QPU and simulating it classically. It is not a speed-up over classical PDE solvers, which would likely be both more accurate and faster on these problems. Whether QRN forecasts can scale to a point where they can be useful for warm-starting classical solvers remains open and is left to future work.

\begin{table}[t]
  \centering
  \footnotesize
  \setlength{\tabcolsep}{2pt}
  \renewcommand{\arraystretch}{1.1}
  \begin{tabular}{@{}llcccc@{}}
  \toprule
  Task & Method & Qubits & $K$ & NRMSE$\downarrow$ & $\Delta$(\%)$\uparrow$ \\
  \midrule
  \multirow{3}{*}{\begin{tabular}[c]{@{}l@{}}Cylinder\\\textit{one-step}
  % \\$z_t \mapsto \hat{z}_{t+1}$
  \end{tabular}}
    & Classical Linear     & --- & --- & 0.121          & ---  \\
    & QRN, CUDA-Q         & 26  & 4   & \textbf{0.069} & \textbf{42.8} \\
    & QRN, IBM QPU        & 26  & 2   & 0.081          & 33.2 \\
  \midrule 
  \multirow{3}{*}{\begin{tabular}[c]{@{}l@{}}Cylinder\\\textit{observer}
  % \\$z_t\mapsto \hat{y}_{t+1}$
  \end{tabular}}
    & Classical Linear     & --- & --- & 0.374          & ---  \\
    & QRN, CUDA-Q         & 26  & 4   & \textbf{0.223} & \textbf{40.3} \\
    & QRN, IBM QPU        & 26  & 2   & 0.295          & 21.2 \\
  \midrule 
  \multirow{4}{*}{\begin{tabular}[c]{@{}l@{}}Gray--Scott\\\textit{one-step}\end{tabular}}
    & Classical Linear     & --- & --- & 0.225          & ---  \\
    & QRN, CUDA-Q         & 26  & 1   & \textbf{0.133} & \textbf{41.1} \\
    & QRN, IBM QPU        & 26  & 1   & \textbf{0.133} & \textbf{41.1} \\
    & QRN, IBM QPU        & 52  & 1   & \textbf{0.133} & \textbf{41.1} \\
  \bottomrule
  \end{tabular}
  \caption{One-step forecasting and observer prediction errors.
  One-step targets are cylinder velocity or Gray--Scott species
  concentrations; the observer target is cylinder pressure at $t+1$
  from velocity at $t$. The linear baseline is the same ridge readout
  on the GAE latent state with the QRN features replaced by classical linear regression,
  $\hat z_{t+1}=Wz_t+b$. $K$ is the maximum retained Pauli weight
  (all nonidentity Pauli-$Z$ strings of weight $\le K$).
  $\Delta$ is the reduction in NRMSE relative to the baseline for the
  same task. IBM QPU results are from \texttt{ibm\_boston}.
  Best result per task in bold.}
  \label{tab:summary}
  \end{table}

\begin{table}[t]
\centering
\small
\setlength{\tabcolsep}{3.5pt}
\renewcommand{\arraystretch}{1.08}
\begin{tabular}{@{}lcccc@{}}
\toprule
Task & Qubits & QPU (s) & H100 (s) & Speed-up \\
\midrule
\multicolumn{5}{@{}l}{\emph{Cylinder Flow}}\\
one-step    & 26 & 75  & $2.5\times10^5$ & ${\sim}3300\times$ \\
closed-loop & 26 & 465 & $8.5\times10^5$ & ${\sim}1800\times$ \\
\midrule
\multicolumn{5}{@{}l}{\emph{Gray--Scott}}\\
one-step    & 26 & 264 & $8.5\times10^5$ & ${\sim}3200\times$ \\
one-step    & 52 & 222 & --- & --- \\
closed-loop & 52 & 320 & --- & --- \\
\bottomrule
\end{tabular}
\caption{
Execution time per experiment for the QRN circuits, normalized to 30{,}000 shots. QPU entries are \texttt{ibm\_boston} device execution times, excluding queueing. GPU entries are CUDA-Q exact statevector sampling of the same circuits on a single NVIDIA H100, with parallel shot jobs summed. Closed-loop rows are complete 10-step rollouts, whose steps execute strictly in sequence. The 52-qubit circuits are beyond exact statevector simulation (---).
}
\label{tab:runtimes}
\end{table}

\section{Conclusion}
In this letter, we realize unstructured nonlinear PDE forecasting with a recurrent QRN on IBM QPUs and in CUDA-Q simulation. A graph autoencoder compresses PDE fields on arbitrary meshes into latent vectors, and a fixed quantum circuit with partial measure-and-reset generates features for a readout that is classically regressed against the data, so no circuit parameter is trained. A dynamic observable reconstruction method lets the Pauli weight $K$ be chosen after measurement to match the shot budget and hardware noise, and a sliding-window protocol enables closed-loop autoregressive PDE forecasting on a QPU.

The central finding is that a very large, untrained quantum kernel space adds predictive value that survives hardware noise. Despite the reservoir evolving on a state space of up to $2^{52}$ dimensions (well-beyond the memory capacity of exact classical simulation) and on circuits of lengths up to ${\sim}28\times T_1$, its features reduce one-step error by 21--43\% relative to the classical architecture components alone. Noise degrades one-step accuracy only modestly and, in closed-loop rollouts, actually improves it, which is consistent with noise acting as a regularizer. While we again emphasize that this work makes no claim of practical advantage, which would require ablation studies against the best classical methods, it does  present evidence that large, gradient-free quantum kernel spaces can remain useful in the presence of noise, and even benefit from it. These findings emphasize that recurrent QRC is a promising direction for further exploration.

\par\addvspace{0.75\baselineskip}
\begin{acknowledgments}
\noindent\textit{Acknowledgments ---}
We acknowledge helpful and motivating conversations with Ian Pegler at NVIDIA, which motivated the experiments on predicting pressure given velocity. We acknowledge the use of IBM Quantum Credits through the IBM Quantum Startups Program. This work used Bridges-2 at the Pittsburgh Supercomputing Center through allocation CIS260063 from the Advanced Cyberinfrastructure Coordination Ecosystem: Services \& Support (ACCESS) program, supported by National Science Foundation Grants No.~2138259, No.~2138286, No.~2138307, No.~2137603, and No.~2138296. The author EC acknowledges support from the DoW SMART (Science, Mathematics, and Research for Transformation) Scholarship. The views expressed in this paper are those of the authors and do not reflect the official policy or position of the aforementioned parties.
\end{acknowledgments}

\par\addvspace{0.5\baselineskip}
\noindent\textit{Code availability ---}
The code implementing the QRN model used in this Letter is
available at \url{https://github.com/econnerty/QRC-partial-SWAP}.

\par\addvspace{0.5\baselineskip}
\noindent\textit{Author Contributions ---}
E.C. conducted experiments and helped design the architecture. E.N.E. suggested the application area, helped design the architecture, and provided feedback. J.E. provided data for the experiments. T.L.P. provided project and editorial feedback and helped interpret the results. G.A. helped with ideation and editorial feedback.

%\begin{acknowledgments}
%\end{acknowledgments}

\bibliography{sn-bibliography-2}

\clearpage
\onecolumngrid

% Embedded Supplemental Information
\begin{center}
\rule{\linewidth}{4pt}\par
\vspace{0.3cm}
{\Large\bfseries Realizing Unstructured PDE Forecasting \\ on Superconducting Quantum Processors\par}
\vspace{0.3cm}
{Erik Connerty\textsuperscript{1,2,$\dagger$},
Jonathan Eastridge\textsuperscript{3}, Taylor Patti\textsuperscript{4}, Gerasimos Angelatos\textsuperscript{5},
Ethan N. Evans\textsuperscript{1}\par}
\smallskip
{\small \textsuperscript{1}QodeX Quantum, Chicago, IL, USA\\
\textsuperscript{2}AI Institute, University of South Carolina, Columbia, SC, USA\\
\textsuperscript{3}Naval Surface Warfare Center Panama City Division, Panama City, FL, USA\\
\textsuperscript{4}NVIDIA, Santa Clara, CA, USA\\
\textsuperscript{5}RTX BBN, Cambridge, MA, USA\par}
\vspace{0.3cm}
\rule{\linewidth}{1.5pt}
\end{center}

\appendix
\setcounter{figure}{0}
\renewcommand{\thefigure}{S\arabic{figure}}
\setcounter{table}{0}
\renewcommand{\thetable}{S\arabic{table}}
\setcounter{equation}{0}
\renewcommand{\theequation}{S\arabic{equation}}
\providecommand{\theHfigure}{}\renewcommand{\theHfigure}{S\arabic{figure}}
\providecommand{\theHtable}{}\renewcommand{\theHtable}{S\arabic{table}}
\providecommand{\theHequation}{}\renewcommand{\theHequation}{S\arabic{equation}}
\providecommand{\theHalgorithm}{}\renewcommand{\theHalgorithm}{S\arabic{algorithm}}
\section*{Supplementary Material}

\suppnote{si:cylinder}{2D cylinder flow simulation details}
Data for the 2D cylinder system is generated by solving the
incompressible unsteady Reynolds-averaged Navier--Stokes (URANS)
equations for the mean velocity $\bar u_i$ and mean pressure
$\bar p$,
\begin{align}
\frac{\partial \bar u_i}{\partial x_i} &= 0, \label{eq:urans-continuity}\\
\frac{\partial \bar u_i}{\partial t}
  + \bar u_j \frac{\partial \bar u_i}{\partial x_j}
  &= -\frac{1}{\rho}\frac{\partial \bar p}{\partial x_i}
   + \frac{\partial}{\partial x_j}\!\left[(\nu + \nu_t)
     \left(\frac{\partial \bar u_i}{\partial x_j}
     + \frac{\partial \bar u_j}{\partial x_i}\right)\right],
   \label{eq:urans-momentum}
\end{align}
in which the Reynolds stresses are represented through the Boussinesq
hypothesis by an eddy viscosity $\nu_t$. Closure is provided by the
$k$--$\omega$ SST model \cite{Menter_1994,Menter_2003}, which transports the
turbulent kinetic
energy $k$ and the specific dissipation rate $\omega$,
\begin{align}
\frac{\partial k}{\partial t}
  + \bar u_j \frac{\partial k}{\partial x_j}
  &= P_k - \beta^{*} k \omega
   + \frac{\partial}{\partial x_j}\!\left[(\nu + \sigma_k \nu_t)
     \frac{\partial k}{\partial x_j}\right],
   \label{eq:sst-k}\\
\frac{\partial \omega}{\partial t}
  + \bar u_j \frac{\partial \omega}{\partial x_j}
  &= \frac{\gamma}{\nu_t} P_k - \beta \omega^{2}
   + \frac{\partial}{\partial x_j}\!\left[(\nu + \sigma_\omega \nu_t)
     \frac{\partial \omega}{\partial x_j}\right] \nonumber\\
  &\quad + 2(1 - F_1)\,\frac{\sigma_{\omega 2}}{\omega}\,
     \frac{\partial k}{\partial x_j}
     \frac{\partial \omega}{\partial x_j},
   \label{eq:sst-omega}
\end{align}
where $P_k$ is the turbulence production term, $F_1$ the SST blending
function that recovers $k$--$\omega$ near walls and $k$--$\epsilon$ in
the free stream, and the eddy viscosity carries the shear-stress
limiter $\nu_t = a_1 k / \max(a_1 \omega,\, S F_2)$ with strain-rate
magnitude $S$ and second blending function $F_2$. The model constants
take their standard values according to the updates provided in
\cite{Menter_2003}. The Reynolds number based on the cylinder
diameter is $\mathrm{Re}_D = U_\infty D / \nu = 1000$.

The system is integrated
with OpenFOAM v2512 via the finite-volume solver
\texttt{pisoFoam} with two pressure-correction loops at a fixed time
step of $\Delta t = \num{e-3}$~s. Temporal discretization is
first-order implicit (backward Euler). Spatial discretization is
second order for gradient and Laplacian terms, the latter with
explicit non-orthogonal correction matching the corrected
surface-normal-gradient scheme, with blended first/second-order
schemes for the divergence terms (\texttt{linearUpwind} for momentum,
\texttt{limitedLinear} for the turbulence quantities).

The computational domain spans $20D \times 10D$ with the cylinder
centered $4D$ downstream of the inlet, meshed by
\texttt{snappyHexMesh} into a hex-dominant grid of 33{,}824 cells
(33{,}004 hexahedra, 72 prisms, and 748 polyhedra). The near-wall
resolution is sufficient to dispense with wall functions for the
turbulence variables: the maximum $y^+$ on the cylinder wall is
approximately 0.048 throughout the simulation.

Boundary conditions are a uniform Dirichlet inlet [$U = (2, 0)$~m/s,
$k = \num{6e-4}$~m$^2$/s$^2$, $\omega = 0.049$~s$^{-1}$] with
zero-gradient pressure; a fixed-pressure outlet ($p = 0$) with
zero-gradient $U$, $k$, and $\omega$; zero-gradient top and bottom
boundaries; and a no-slip cylinder wall ($U = \mathbf{0}$,
$k = \omega = 0$) with zero-gradient pressure. From this simulation,
300 snapshots of $(U_x, U_y)$ are retained at 0.1~s intervals
(every 100 solver steps) in the periodic vortex-shedding regime. The
pressure field $p$ of the same simulation supplies the observer target
of Supplementary Note~\ref{si:pressure}.

\suppnote{si:grayscott}{Gray--Scott reaction--diffusion system}
Gray--Scott data is obtained from the
\texttt{gray\_scott\_reaction\_diffusion} collection of The Well
benchmark suite \cite{NEURIPS2024_4f9a5acd}, and realizes the
reaction--diffusion model of Gray and Scott in the form studied by
Pearson \cite{pearsongrayscott}. Two chemical species with
concentrations $A(\mathbf{x}, t)$ and $B(\mathbf{x}, t)$ evolve under
the coupled reaction--diffusion equations
\begin{align}
\frac{\partial A}{\partial t}
  &= \delta_A \Delta A - AB^{2} + f\,(1 - A), \label{eq:gs-a}\\
\frac{\partial B}{\partial t}
  &= \delta_B \Delta B + AB^{2} - (f + k)\,B, \label{eq:gs-b}
\end{align}
where $\Delta$ is the Laplacian, $\delta_A$ and $\delta_B$ are the
species diffusivities, $f$ is the rate at which $A$ is fed into the
domain, and $k$ is the rate at which $B$ is removed. The cubic
autocatalytic term $AB^{2}$ implements the reaction
$A + 2B \rightarrow 3B$, consuming $A$ and producing $B$. The
competition between this autocatalysis and the linear feed and kill
terms generates the stiff, localized kinetics that make the system a
complementary test to the advection-dominated cylinder flow.

The trajectory used here has diffusivities
$\delta_A = \num{2.0e-5}$ and $\delta_B = \num{1.0e-5}$ on the
periodic domain $[-1, 1]^2$ discretized at $128 \times 128$, with the
\texttt{spirals} parameter set ($f = 0.018$, $k = 0.051$;
validation trajectory~3). The 1{,}001 retained snapshots span
$t \in [0, 10^4]$ at intervals of 10 time units.

\suppnote{si:autoencoderdetails}{Autoencoders}
Table~\ref{tab:autoencoders} records the architectures used for the field autoencoders.
Cylinder velocity $|U|$ drives the QRN using 13 latents; the separate 48 pressure latents are observer targets only.
Gray--Scott uses a convolutional autoencoder with 26 latents. The decoded fields include both errors from the QRN inference and inherent autoencoder reconstruction error.

The 2D cylinder graph encoders use the solver's owner--neighbor cell adjacency,
so compression preserves the unstructured mesh connectivity. Their
coordinate-conditioned decoders reconstruct fields at the original cell
centres. The pressure encoder is trained separately, with an additional
mesh-edge gradient loss to retain spatial pressure differences. For the
uniform Gray--Scott grid, residual convolutional layers provide the spatial
encoder and transposed convolutions reconstruct both species. The same
trained decoders are used for one-step and closed-loop predictions. Different architectures were used as seen fit to improve reconstruction accuracy.

\begin{table}[!ht]
\centering
% \scriptsize
\renewcommand{\arraystretch}{1.15}
\begin{tabular}{@{}llll@{}}
\toprule
 & \textbf{2D cyl.\ velocity} & \textbf{2D cyl.\ pressure} & \textbf{Gray--Scott} \\
 & Graph AE & Graph AE & Conv.\ AE \\
\midrule
Input & \shortstack[l]{$(x,y,U_x,U_y)$ on\\33{,}824 mesh cells} & \shortstack[l]{$(x,y,p)$ on\\33{,}824 mesh cells} & \shortstack[l]{$A,B$ on\\$128\times128$ grid} \\
\addlinespace[2pt]
Encoder & \shortstack[l]{4 residual\\GraphSAGE layers} & \shortstack[l]{3 residual\\GraphSAGE layers} & \shortstack[l]{4 stride-2 conv.\ layers\\+ residual blocks} \\
\addlinespace[2pt]
Latent & $d_z = 13$ ($\tanh$) & $d_z = 48$ ($\tanh$) & $d_z = 26$ ($\tanh$) \\
\addlinespace[2pt]
Decoder & \shortstack[l]{Coordinate MLP +\\2 GraphSAGE layers} & \shortstack[l]{Coordinate MLP +\\2 GraphSAGE layers} & Transposed convolutions \\
\addlinespace[2pt]
Parameters & $4.1 \times 10^{5}$ & $4.1 \times 10^{5}$ & $2.5 \times 10^{6}$ \\
\addlinespace[2pt]
Test MSE & $1.6 \times 10^{-2}$ & $1.9 \times 10^{-3}$ & $2.1 \times 10^{-3}$ \\
\addlinespace[2pt]
Split & 240 / 30 / 30 & 210 / 60 / 30 & 701 / 150 / 150 \\
\bottomrule
\end{tabular}
\caption{Autoencoder architectures. The cylinder encoders operate on
the OpenFOAM owner--neighbour cell-adjacency graph, and their implicit
decoders (Fourier-feature coordinate MLPs) reconstruct all cell
centres. Global readout is attention + max pooling. The pressure
autoencoder provides the 48-dimensional observer target of
Supplementary Note~\ref{si:pressure}. It is trained with an added
mesh-edge pressure-gradient loss (weight $0.1$) that penalizes
mismatched pressure differences between neighbouring cells. Test MSE
is on normalized fields, and Split is the random train/val/test snapshot
partition (the pressure model uses a $70/20/10$ shuffle, seed
$\eta = 2127$).}
\label{tab:autoencoders}
\end{table}

\suppnote{si:pressure}{Pressure-field observer}
The pressure field from the 2D cylinder data is predicted using an ``observer'' style inference model where only velocity fields $|U|$ are provided to the model, and the network is tasked with predicting pressure fields. The velocity-driven reservoir measurements are reused to predict 48 pressure latents at $t+1$ with the ``skip-connection'' described in the main manuscript. This additional experiment requires no additional quantum circuits executions since observables are re-used and is generally considered more difficult than a standard one-step inference task. Pressure and velocity errors use different target variances for
normalization, so their NRMSE values should be interpreted relative to their
respective baselines.

\begin{figure}[!ht]
  \centering
  \includegraphics[width=0.72\textwidth]{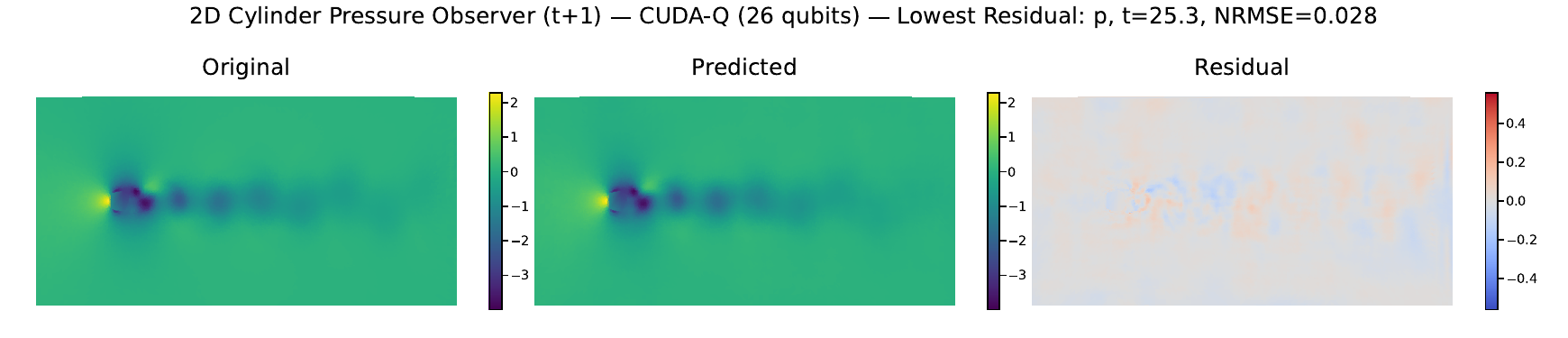}\\[2pt]
  \includegraphics[width=0.72\textwidth]{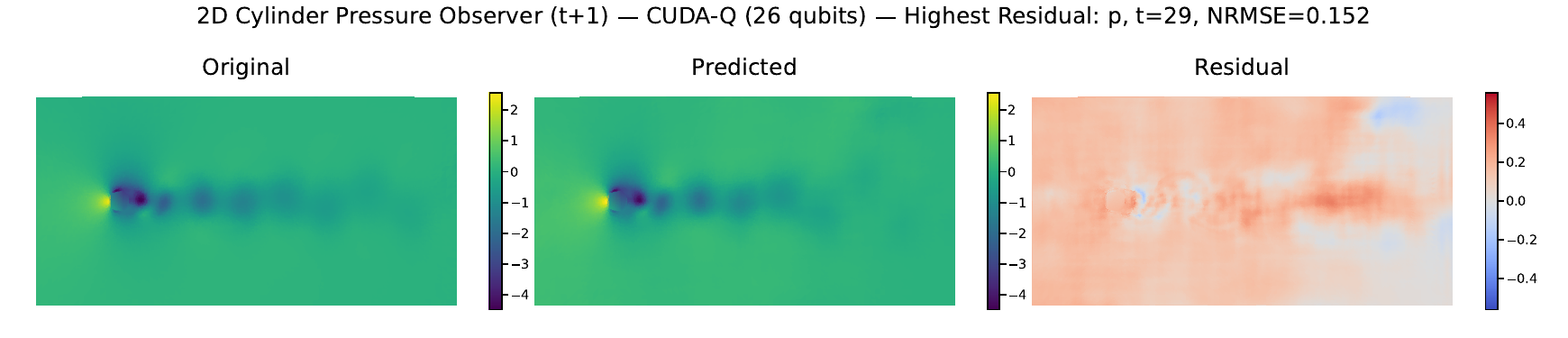}
  \caption{CUDA-Q pressure observer: lowest- (top) and highest-residual (bottom) test frames, showing truth, prediction, and residual. Field NRMSE includes autoencoder error.}
  \label{fig:pressure-cudaq-decoded}
\end{figure}

\begin{figure}[!ht]
  \centering
  \includegraphics[width=0.72\textwidth]{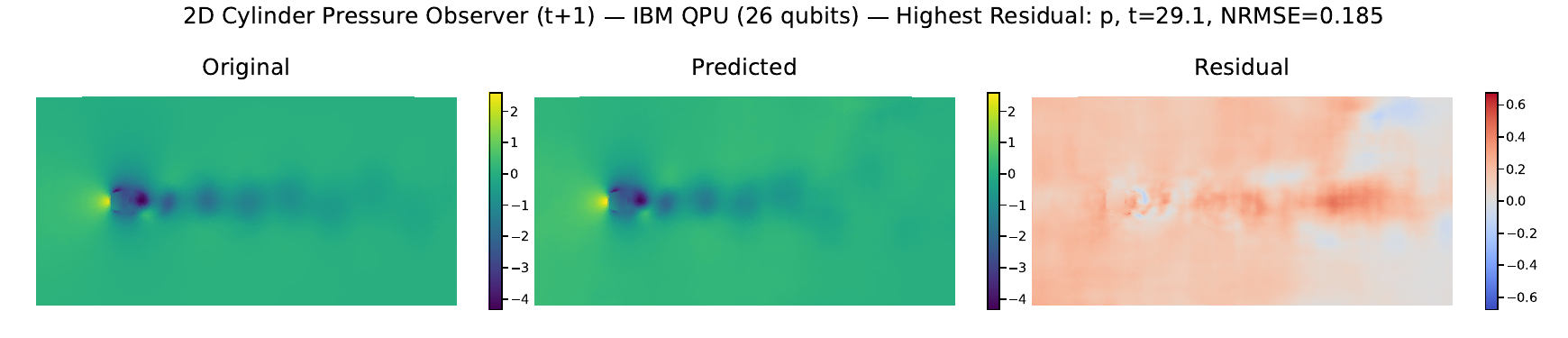}
  \caption{IBM pressure observer: highest-residual test frame, complementing the representative pressure reconstruction in the Letter. Field NRMSE includes autoencoder error.}
  \label{fig:pressure-ibm-decoded}
\end{figure}

Table~\ref{tab:pressure_params} gives the experiment details and the decoded fields are shown in Figs.~\ref{fig:pressure-cudaq-decoded},~\ref{fig:pressure-ibm-decoded}.
Field NRMSE ranges over the 60 test frames are 0.028--0.152 (CUDA-Q) and 0.030--0.185 (IBM).
Horizon dependence and pressure-latent traces appear in Notes~\ref{si:horizons} and~\ref{si:traces}.
\begin{table}[!ht]
    \centering
    % \scriptsize
    \begin{tabular}{llll}
    \toprule
    Name & Parameter & CUDA-Q & IBM QPU \\
    \midrule
    Observable Order & $K$ & 1--4 & 2 \\
    \# of Observable Features & $F$ & 1{,}092 & 91 \\
    Design Matrix Columns & $F + d_z$ & 1{,}105 & 104 \\
    Target Dimension (pressure) & $d_z^{(p)}$ & 48 & 48 \\
    \# of Training Data Points & $n_{\text{train}}$ & 119 & 119 \\
    \# of Test Data Points & $n_{\text{test}}$ & 60 & 60 \\
    \# of Washout Points & $n_{\text{washout}}$ & 120 & 120 \\
    Readout (elastic net) & $\alpha$ & \num{1e-3} & \num{3.2e-2} \\
    Elastic-Net Mixing & $l_1$ & 0.1 & 0.1 \\
    \bottomrule
    \end{tabular}
    \caption{2D Cylinder Pressure-Observer ($t{+}1$) Parameters. The
    reservoirs, shot budgets, and recorded observables are identical
    to the featured velocity runs
    (Table~\ref{tab:teacher-params});
    only the readout differs. Predictions are clipped to $[-1,1]$;
    the readout is selected on $t{+}1$ test RMSE over the standard
    88-configuration grid.}
    \label{tab:pressure_params}
\end{table}


\suppnote{si:qrn}{Quantum reservoir network hyperparameter sweep}

\begin{figure}[!htbp]
\centering
\includegraphics[width=0.85\linewidth,trim=0 188 0 0,clip]{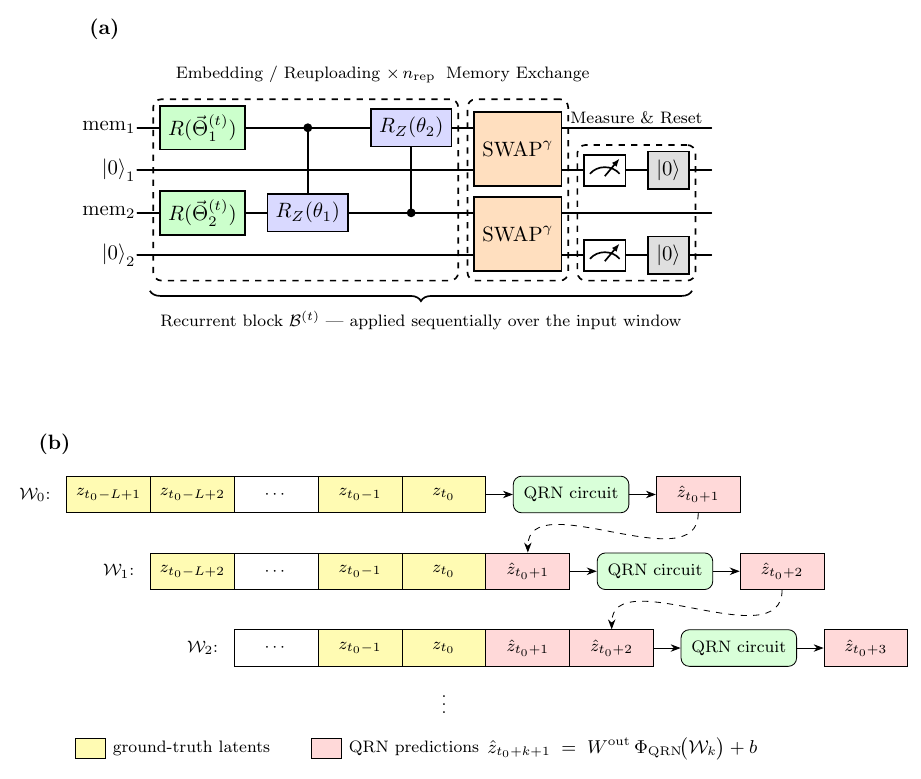}
\caption{The gate-level QRN schematic. Input rotations and memory-register entangling blocks comprise the Embedding/Reuploading later, which is repeated $n_{rep}$ times. The hyperparameter $\gamma$ controls the partial-SWAP exchange in the Memory Exchange layer. Afterwards, readout qubits are measured and then reset.}
\label{fig:si-circuit}
\end{figure}

In this section, we examine the effects of amplitude-damping
strength $\gamma$ and the number of reuploading blocks detailed in Fig.~\ref{fig:si-circuit}, as well as the maximum
Pauli weight $K$ included in the readout features. These parameters
control memory capacity, circuit depth, and the inclusion of
higher-order correlations through Pauli weight, respectively. 
Table~\ref{tab:teacher-params} records the various QRN configurations used
in the main manuscript. 

\begin{table}[!htbp]
\centering
% \footnotesize
\begin{tabular}{llrrrrrl}
\toprule
System/platform & Qubits & $\gamma$ & Blocks & $K$ & Shots & Seed & Readout $(\alpha,l_1)$ \\
\midrule
Cylinder/CUDA-Q & 26 & .75 & 3 & 4 & 30,000 & 123123 & EN $(10^{-3},.25)$ \\
Cylinder/IBM & 26 & .75 & 2 & 2 & 30,000 & 27183 & Lasso $(7.5\!\times\!10^{-4},-)$ \\
Gray--Scott/CUDA-Q & 26 & .75 & 3 & 1 & 10,000 & 14021 & Lasso $(1.8\!\times\!10^{-2},-)$ \\
Gray--Scott/IBM & 26 & .75 & 3 & 1 & 30,000 & 14021 & Lasso $(1.3\!\times\!10^{-2},-)$ \\
Gray--Scott/IBM & 52 & .9 & 2 & 1 & 30,000 & 27183 & Lasso $(1.3\!\times\!10^{-2},-)$ \\
\bottomrule
\end{tabular}
\caption{one-step reservoir and selected readout parameters.
EN denotes elastic net; $\alpha$ is the regularization strength and
$l_1$ is the elastic-net mixing ratio. $K$ is the maximum Pauli
weight included in the readout features. The 26 qubit Gray--Scott
hardware run matches the simulation reservoir. IBM shot totals
pool 3, 4, and 6 jobs for cylinder, 26 qubit Gray--Scott, and
52 qubit Gray--Scott, respectively.}
\label{tab:teacher-params}
\end{table}

\begin{figure}[!ht]
  \centering
  \includegraphics[width=0.95\textwidth]{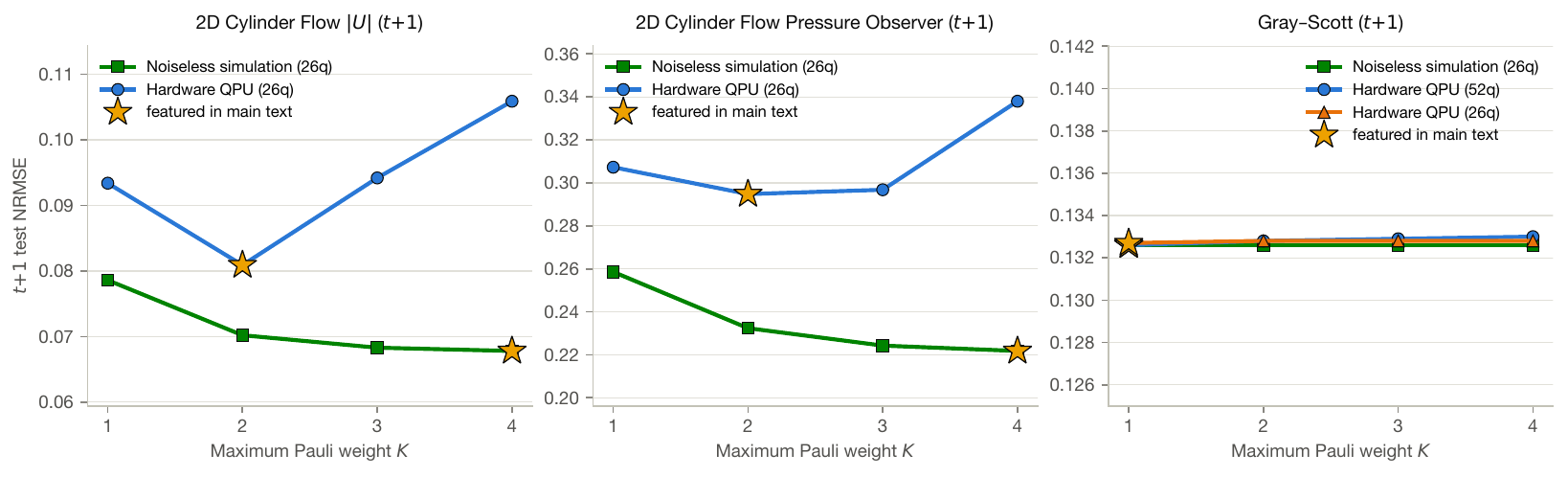}
  \caption{Pauli weight comparison for cylinder velocity (left),
  pressure observer (middle), and Gray--Scott (right). Readouts
  are reselected at each order using the same recorded counts.
  Stars mark the featured choices. Cylinder hardware achieves
  its minimum error at $K=2$, while Gray--Scott changes little
  across orders. Lower is better.}
  \label{fig:order-ablation}
\end{figure}

To test whether higher Pauli-weight observables, and thus more observables,
improve predictions, we provide an ablation over Pauli-weight $K$ in Figure~\ref{fig:order-ablation}. For 13
readout qubits, orders $K=1,2,3,4$ yield 13, 91, 377, and 1,092
features, respectively; for 26 readout qubits, they yield 26, 351,
2,951, and 17,901 features. Higher orders improve the noiseless cylinder
predictions, whereas cylinder hardware reaches its minimum error
at $K=2$ and then degrades, consistent with reduced usefulness of
higher-order correlations under device noise. Gray--Scott accuracy
changes little across orders at the executed shot budgets; whether
additional shots would reveal a benefit remains unresolved as compute times ballooned rapidly for the larger time-series data, limiting shot count in simulation.

\begin{figure}[!ht]
  \centering
  \includegraphics[width=0.9\textwidth]{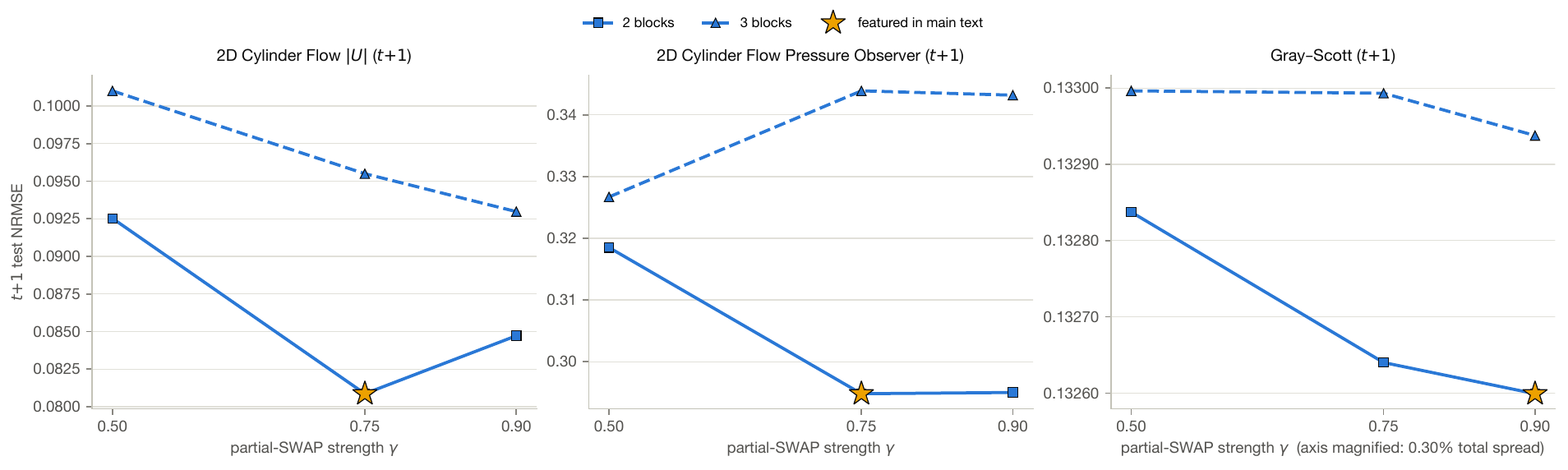}
  \caption{Hardware reservoir sweep at seed 27183: test NRMSE
  versus amplitude-damping strength $\gamma$ for two or three
  reuploading blocks. Panels show cylinder velocity, cylinder
  pressure, and Gray--Scott. Stars mark selected configurations;
  each point represents a single run with its best test-selected
  readout. The Gray--Scott vertical scale is expanded.}
  \label{fig:hyperparam-ablation}
\end{figure}

Figure~\ref{fig:hyperparam-ablation} details various QRN configurations that  perturb damping parameter $\gamma$ and the number of reuploading blocks. We observe stronger sensitivity in the 2D cylinder flow task than for Gray--Scott.
Two reuploading blocks outperform three across the tested cylinder
damping strengths, with the same reservoir selected for velocity
and pressure prediction. Gray--Scott varies by only about 0.3\%
across the sweep.

\suppnote{si:arproto}{Closed-loop rollouts and window comparisons}
We provide further detail on the autoregressive rollouts in this section. For each predicted step, a window $L$ is used to re-initialize the reservoir by replaying the current latent window. Only the final time-step of observables provide features to the learned readout. The prediction of the readout is then appended and the oldest latent removed.
All rollouts use order-1 observables to limit noisy higher-order features being fed back.
The 2D cylinder uses $L=100$ and targets 240--249; Gray--Scott uses $L=40$ and targets 501--510. All reported rollout metrics cover 10 steps.

The parameters for these experiments are given in Table~\ref{tab:rollout-params}. For the 52 qubit Gray--Scott rollout, a three reuploading block reservoir is trained and used in contrast to the two block circuit from the main manuscript, as it was observed to perform better.

\begin{figure}[!htb]
  \centering
  \includegraphics[width=\linewidth]{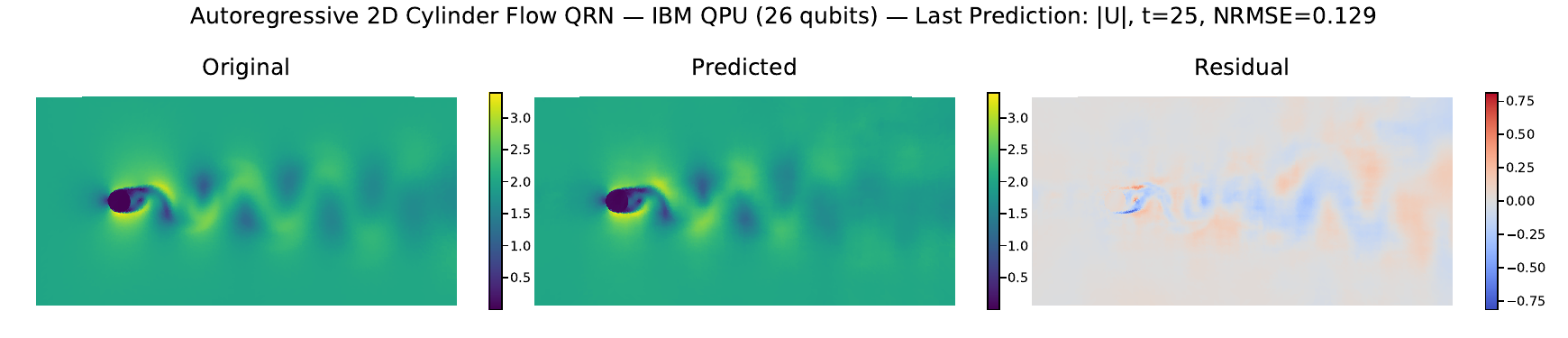}
  \caption{Decoded 2D Cylinder field from the final step of the hardware
  autoregressive rollout (truth / prediction / residual), for the
  10-step velocity-magnitude forecast. The quoted NRMSE is computed against the original (pre-encoding) field and therefore includes the autoencoder reconstruction error.}
  \label{fig:ar-decoded}
\end{figure}

\begin{figure}[!htb]
  \centering
  \includegraphics[width=\linewidth]{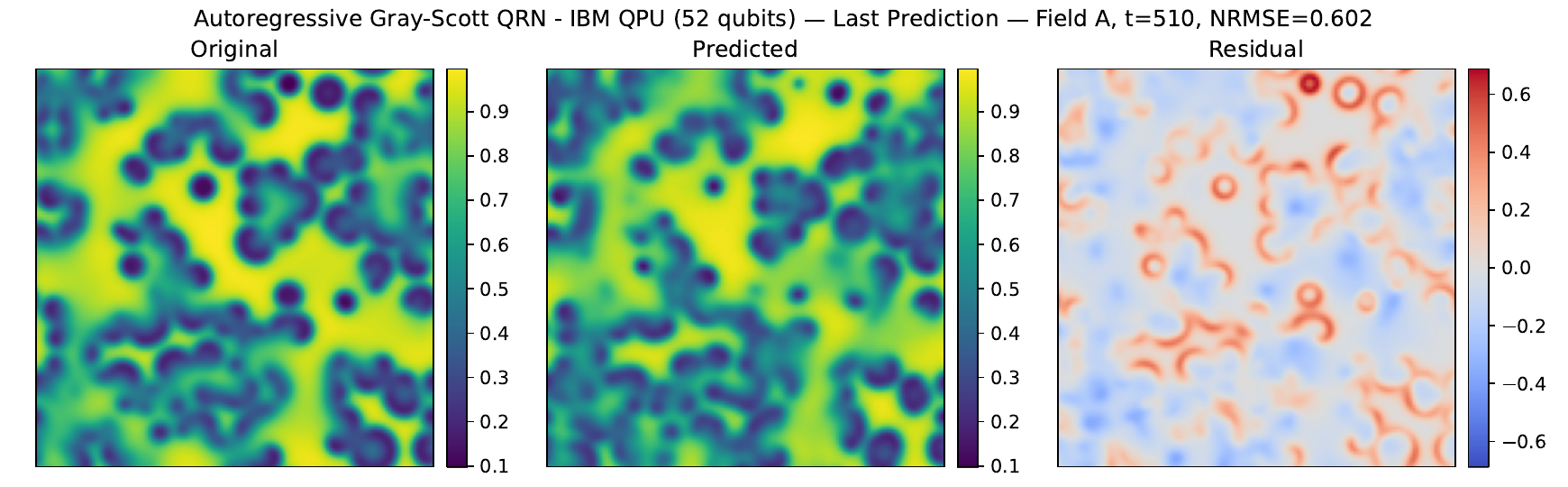}\\[2pt]
  \includegraphics[width=\linewidth]{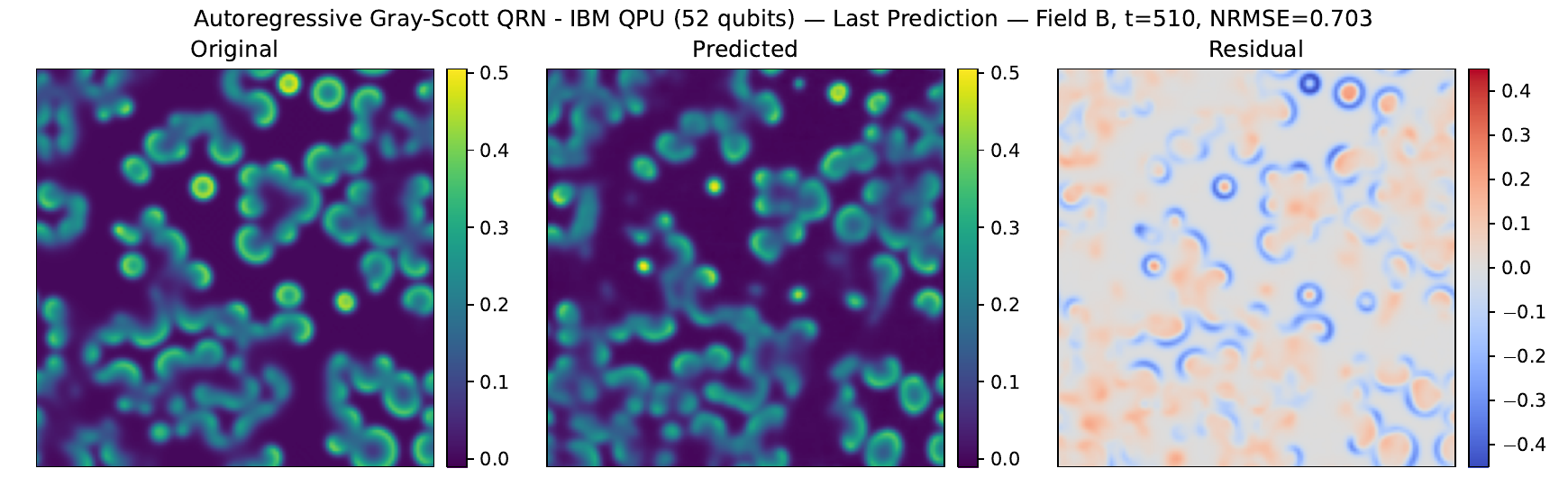}
  \caption{Decoded Gray--Scott species $A$ (top) and $B$ (bottom) at
  the final step of the 52 qubit hardware autoregressive rollout
  (truth / prediction / residual). The quoted NRMSE is computed against the original (pre-encoding) field and therefore includes the autoencoder reconstruction error.}
  \label{fig:gs-ar-decoded}
\end{figure}

Table~\ref{tab:ar-window-ablation} records an ablation over several window choice $L$.
A longer replay can reduce initialization effects, but the observed score
also depends on the learned readout and on errors already fed back into the
window. The selected cylinder deployment uses the longest tested window ($L=100$),
whereas Gray--Scott performs best here at $L=40$. These comparisons do not
identify a universal optimal window length. The final-step decoded fields
are shown in Figs.~\ref{fig:ar-decoded} and~\ref{fig:gs-ar-decoded};
Note~\ref{si:traces} resolves their evolution in latent space.

Notably, the featured cylinder hardware rollout of Fig.~\ref{fig:ar-noise-ablation} of the main manuscript retains the final-block CRZ layer omitted in training: its NRMSE is 0.257, versus 0.318 when that layer is omitted in rollout too. This increase in performance is thought to be due to the observation that increased training noise actually helps autoregressive rollout, but warrants further investigation.
CUDA-Q and IBM cylinder rollouts differ in seed and shots, so the platform comparison is not a controlled noise ablation. The authors note that the differences in configurations cannot explain the performance difference alone, as the simulated circuits generally perform better in one-step forecasting but not in the autoregressive rollouts.
The CUDA-Q noise-injection variants in the Letter refit the readout using eight Gaussian-noise replicas of the observable columns ($\sigma=0.03,0.06$), selecting ridge with $\alpha=10^{-6}$.

\begin{table}[!htbp]\centering
\begin{tabular}{llrrrrll}\toprule
System/platform & Qubits & $\gamma$ & $L$ & Shots & Seed & EN $(\alpha,l_1)$ & NRMSE \\\midrule
Cylinder/CUDA-Q & 26 & .75 & 100 & 10,000 & 123123 & $(10^{-3},.1)$ & .707 \\
Cylinder/IBM & 26 & .75 & 100 & 20,000 & 14021 & $(5.6\!\times\!10^{-3},.25)$ & .257 \\
Gray--Scott/IBM & 52 & .9 & 40 & 20,000 & 14021 & $(10^{-2},.5)$ & .470 \\
\bottomrule\end{tabular}
\caption{Featured closed-loop runs: three reuploading blocks, order-1 features, and 10 predictions throughout. Shots are per predicted latent vector. Persistence NRMSE is 1.126 for cylinder and 0.733 for Gray--Scott. The simulation uses a wrapped entangling ring; hardware uses an open chain.}\label{tab:rollout-params}
\end{table}
\begin{table}[!ht]
\centering
% \footnotesize
\begin{tabular}{lllc}
\toprule
System & $L$ & Readout & NRMSE \\
\midrule
2D cylinder & 20  & lasso, $\alpha{=}10^{-2}$ & 0.337 \\
2D cylinder & 40  & ridge, $\alpha{=}10^{-6}$ & 0.402 \\
2D cylinder & 45  & elastic net, $\alpha{=}5.6{\times}10^{-3}$ & 0.402 \\
2D cylinder & 100 & elastic net, $\alpha{=}5.6{\times}10^{-3}$ & 0.318 \\
2D cylinder & 100 & elastic net, $\alpha{=}5.6{\times}10^{-3}$ & 0.257$^\dagger$ \\
\midrule
Gray--Scott & 20  & lasso, $\alpha{=}10^{-2}$ & 0.511 \\
Gray--Scott & 40  & elastic net, $\alpha{=}10^{-2}$ & 0.470$^\dagger$ \\
Gray--Scott & 45  & lasso, $\alpha{=}1.3{\times}10^{-2}$ & 0.531 \\
Gray--Scott & 100 & lasso, $\alpha{=}1.3{\times}10^{-2}$ & 0.534 \\
\bottomrule
\end{tabular}
\caption{Executed hardware window comparisons, each evaluated over 10 targets. Readouts were selected separately, so this is not an isolated ablation of $L$. Daggers identify the featured runs. The two cylinder $L=100$ entries share the readout and differ in the final-block CRZ layer; the featured entry retains it.}
\label{tab:ar-window-ablation}
\end{table}

\begin{figure}[!ht]
  \centering
  \begin{minipage}[t]{0.49\linewidth}\centering
    \includegraphics[width=\linewidth]{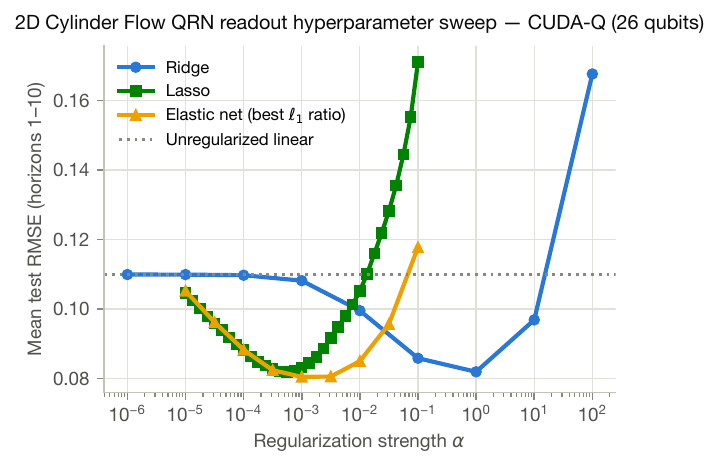}
    \caption{CUDA-Q 2D cylinder: readout hyperparameter sweep (88
    configurations) for the velocity $|U|$ ($t{+}1$) task. The
    pressure-observer readout is swept separately (Supplementary
    Note~\ref{si:pressure}). \emph{Lower is better.}}
    \label{fig:cudaq-sweep}
  \end{minipage}\hfill
  \begin{minipage}[t]{0.49\linewidth}\centering
    \includegraphics[width=\linewidth]{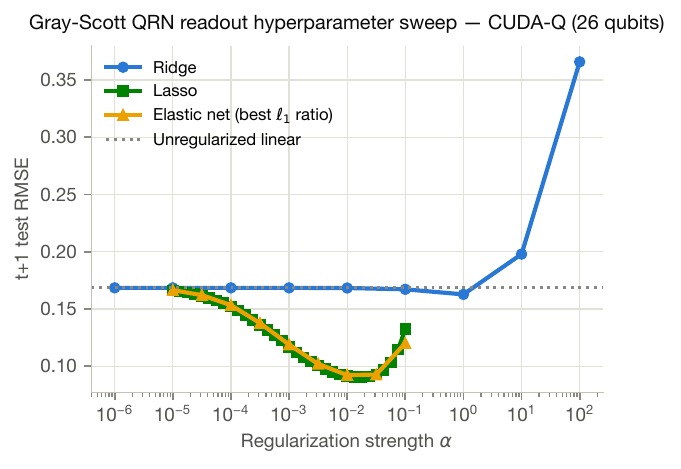}
    \caption{CUDA-Q Gray--Scott: readout hyperparameter sweep for the
    species $A,B$ ($t{+}1$) task. \emph{Lower is better.}}
    \label{fig:cudaq-gs-sweep}
  \end{minipage}
\end{figure}

\begin{figure}[!ht]
  \centering
  \begin{minipage}[t]{0.49\linewidth}\centering
    \includegraphics[width=\linewidth]{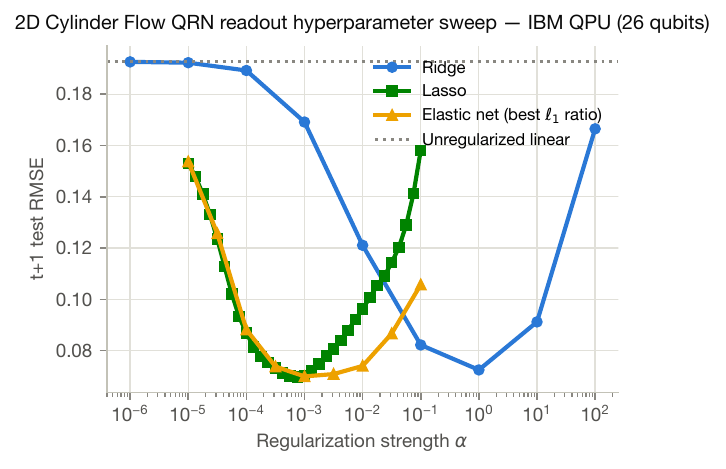}
    \caption{IBM 2D cylinder: readout hyperparameter sweep for the
    velocity $|U|$ ($t{+}1$) task. \emph{Lower
    is better.}}
    \label{fig:ibm2dcyl-sweep}
  \end{minipage}\hfill
  \begin{minipage}[t]{0.49\linewidth}\centering
    \includegraphics[width=\linewidth]{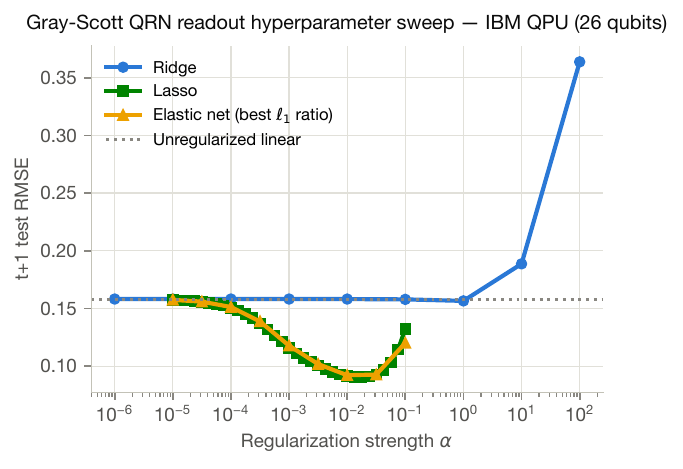}
    \caption{IBM Gray--Scott (26 qubits): readout hyperparameter sweep
    for the species $A,B$ ($t{+}1$) task of the matched-reservoir
    hardware run (Table~\ref{tab:teacher-params}). \emph{Lower is
    better.}}
    \label{fig:gs26-sweep}
  \end{minipage}
\end{figure}

\begin{figure}[!ht]
  \centering
  \begin{minipage}[t]{0.49\linewidth}\centering
    \includegraphics[width=\linewidth]{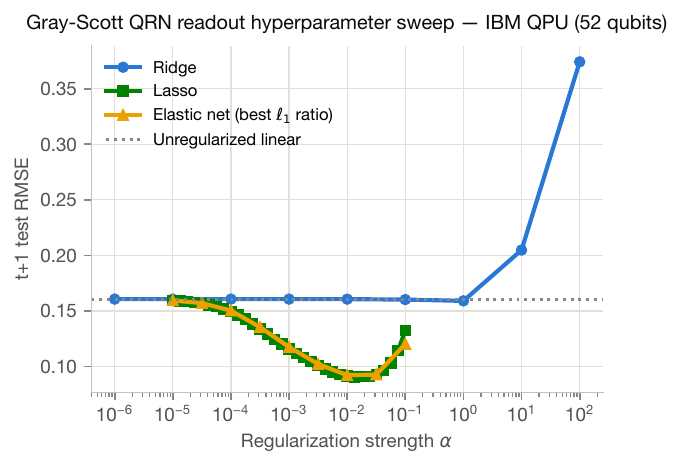}
    \caption{IBM Gray--Scott (52 qubits): readout hyperparameter sweep
    for the species $A,B$ ($t{+}1$) task. \emph{Lower
    is better.}}
    \label{fig:gs-sweep}
  \end{minipage}
\end{figure}

\begin{figure}[!ht]
  \centering
  \begin{minipage}[t]{0.49\linewidth}\centering
    \includegraphics[width=\linewidth]{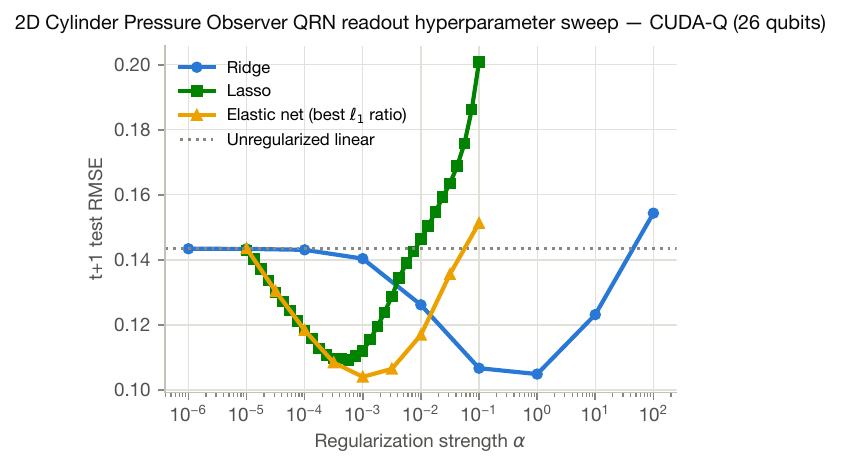}
    \caption{CUDA-Q 2D cylinder pressure-observer ($t{+}1$) task:
    readout hyperparameter sweep (88 configurations). \emph{Lower is
    better.}}
    \label{fig:pressure-cudaq-sweep}
  \end{minipage}\hfill
  \begin{minipage}[t]{0.49\linewidth}\centering
    \includegraphics[width=\linewidth]{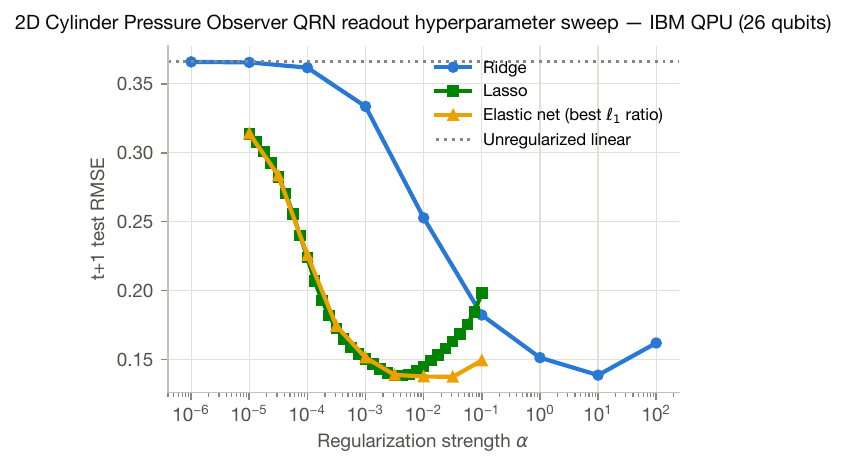}
    \caption{IBM 2D cylinder pressure-observer ($t{+}1$) task: readout
    hyperparameter sweep (88 configurations). \emph{Lower is
    better.}}
    \label{fig:pressure-ibm-sweep}
  \end{minipage}
\end{figure}

\suppnote{si:sweeps}{Readout hyperparameter sweeps}
Here, we provide results from an 88-configuration hyperparameter sweep comparing ridge, lasso, and elastic net across regularization strengths in Figs.~\ref{fig:cudaq-sweep}, \ref{fig:cudaq-gs-sweep}, \ref{fig:ibm2dcyl-sweep}, \ref{fig:gs26-sweep}, \ref{fig:gs-sweep},\ref{fig:pressure-cudaq-sweep}, and \ref{fig:pressure-ibm-sweep}. Each curve shows the family's best test error at each $\alpha$; the unregularized readout is a reference. Pressure sweeps appear in Note~\ref{si:pressure}.

% Ridge penalizes squared coefficients, lasso promotes sparse coefficients,
% and elastic net combines the two penalties. The mixing ratio is optimized
% within the elastic-net family. This regularization is useful when correlated
% observable columns give a feature dimension comparable to or greater than
% the number of training samples.

The IBM and CUDA-Q Gray--Scott forecasting readouts are selected on one-step
test error. The CUDA-Q cylinder velocity readout instead follows mean test
error over horizons $h=1,\ldots,10$, as in the original experiment; this is done to aid the autoregressive task in~Fig.\ref{fig:ar-noise-ablation} by optimizing error over a larger horizon. Pressure
readouts are selected separately on one-step pressure test error.


\begin{figure}[!ht]
  \centering
  \includegraphics[width=0.95\textwidth]{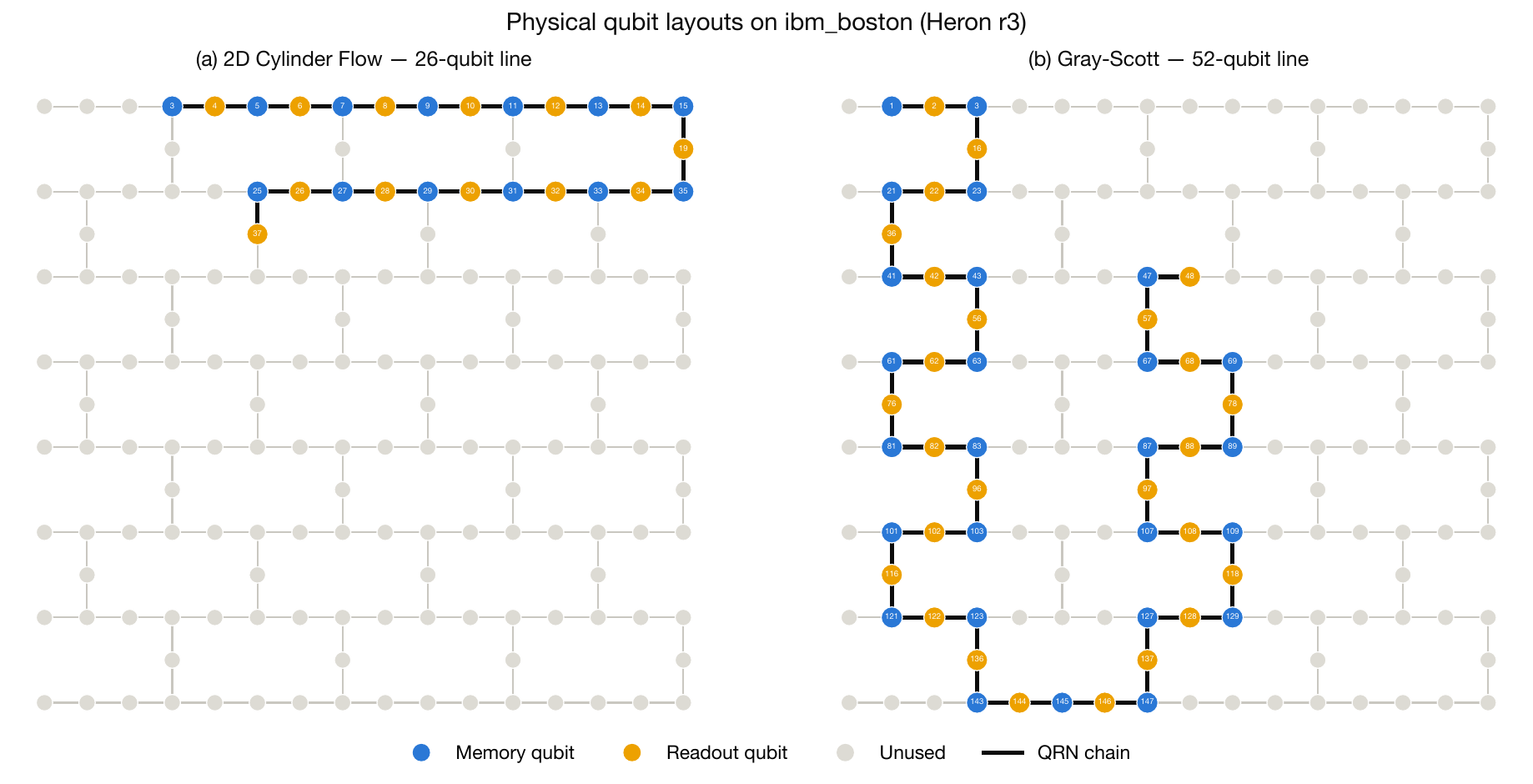}
  \caption{Compiled 26 and 52 qubit layouts on \texttt{ibm\_boston}. Blue and orange mark initial memory and readout assignments; black edges show the chain. Routing can permute assignments within the footprint.}
  \label{fig:qubit-layouts}
\end{figure}

\suppnote{si:layouts}{Physical layouts and compiled circuits}
We provide the layout and various configuration details for the circuits in the main manuscript in this section. Circuit layouts are shown in Fig.~\ref{fig:qubit-layouts} and are extracted from the transpiled training circuits. The compiled statistics distinguish full-trajectory one-step circuits from the autoregressive circuits executed for each rollout step. These details are provided in Table~\ref{tab:compiled-circuits}.

\begin{table}[!ht]
\centering
\small
\begin{tabular}{llccrrrrrr}
\toprule
System & Protocol & Qubits & Reuploading & Steps & Depth & Depth & CZ & CZ & 1q \\
 & & & Blocks & driven & & /step & & /step & gates \\
\midrule
2D cylinder & $t{+}1$        & 26 & 2 & 300   & 14{,}234 & 47.4  & 38{,}670  & 129 & 156{,}068 \\
Gray--Scott & $t{+}1$        & 26 & 3 & 1{,}001 & 99{,}670 & 99.6  & 211{,}921 & 212 & 837{,}891 \\
Gray--Scott & $t{+}1$        & 52 & 2 & 1{,}001 & 53{,}322 & 53.3  & 272{,}827 & 273 & 1{,}081{,}770 \\
2D cylinder & AR (per step)  & 26 & 3 & 100   & 14{,}103 & 141.0 & 28{,}365  & 284 & 106{,}731 \\
Gray--Scott & AR (per step)  & 52 & 3 & 40    & 4{,}466  & 111.7 & 18{,}253  & 456 & 69{,}965 \\
\bottomrule
\end{tabular}
\caption{Compiled circuit statistics of the executed hardware
experiments on \texttt{ibm\_boston}, in the device ISA basis
$\{\mathrm{CZ}, \mathrm{RZ}, \mathrm{SX}, \mathrm{X}\}$. The
one-step ($t{+}1$) rows are the single circuits that drive the
full trajectory. The autoregressive (AR) rows are one rollout step,
which re-drives a window of $L$ time steps and is executed once per
predicted latent vector. ``Steps driven'' is the number of recurrent
time steps in the circuit, and the per-step columns are the totals
divided
by it. One-qubit gates are the RZ, SX, and X totals combined, and each
driven step contributes one measurement and one reset per readout
qubit.}
\label{tab:compiled-circuits}
\end{table}

\begin{figure}[!ht]
  \centering
  \includegraphics[width=0.72\textwidth]{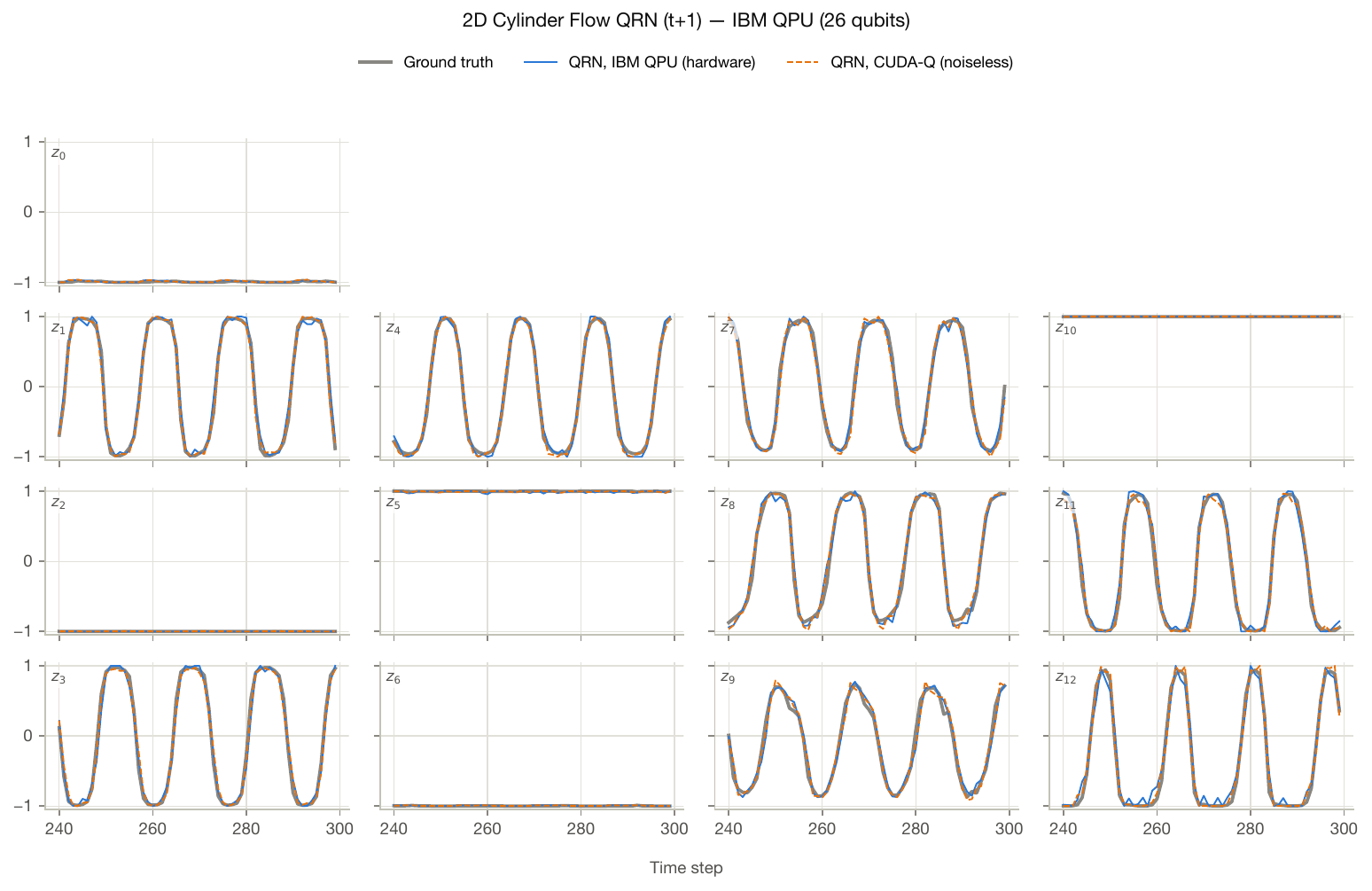}
  \caption{IBM 2D cylinder, velocity $|U|$ ($t{+}1$) task: $h{=}1$
  latent traces on hardware (blue), with the noiseless CUDA-Q
  prediction overlaid (orange, dashed), against the ground truth
  (gray).}
  \label{fig:ibm2dcyl-traces}
\end{figure}

\begin{figure}[!ht]
  \centering
  \includegraphics[width=0.72\textwidth]{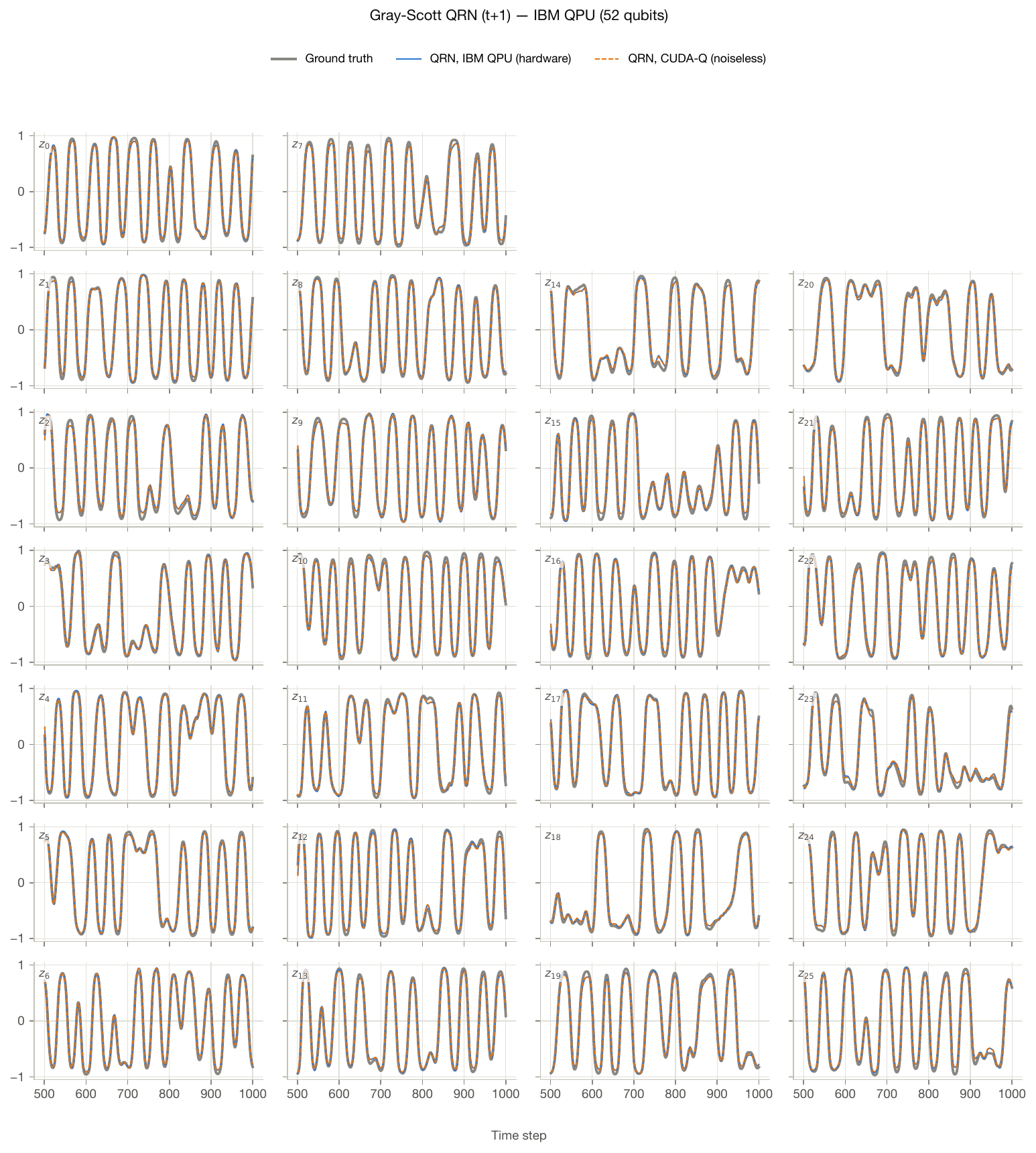}
  \caption{IBM Gray--Scott, species $A,B$ ($t{+}1$) task: $h{=}1$
  latent traces (500-step test window) on hardware (blue), with the
  noiseless CUDA-Q prediction overlaid (orange, dashed), against the
  ground truth (gray).}
  \label{fig:gs-traces}
\end{figure}

\begin{figure}[!ht]
  \centering
  \includegraphics[width=\textwidth]{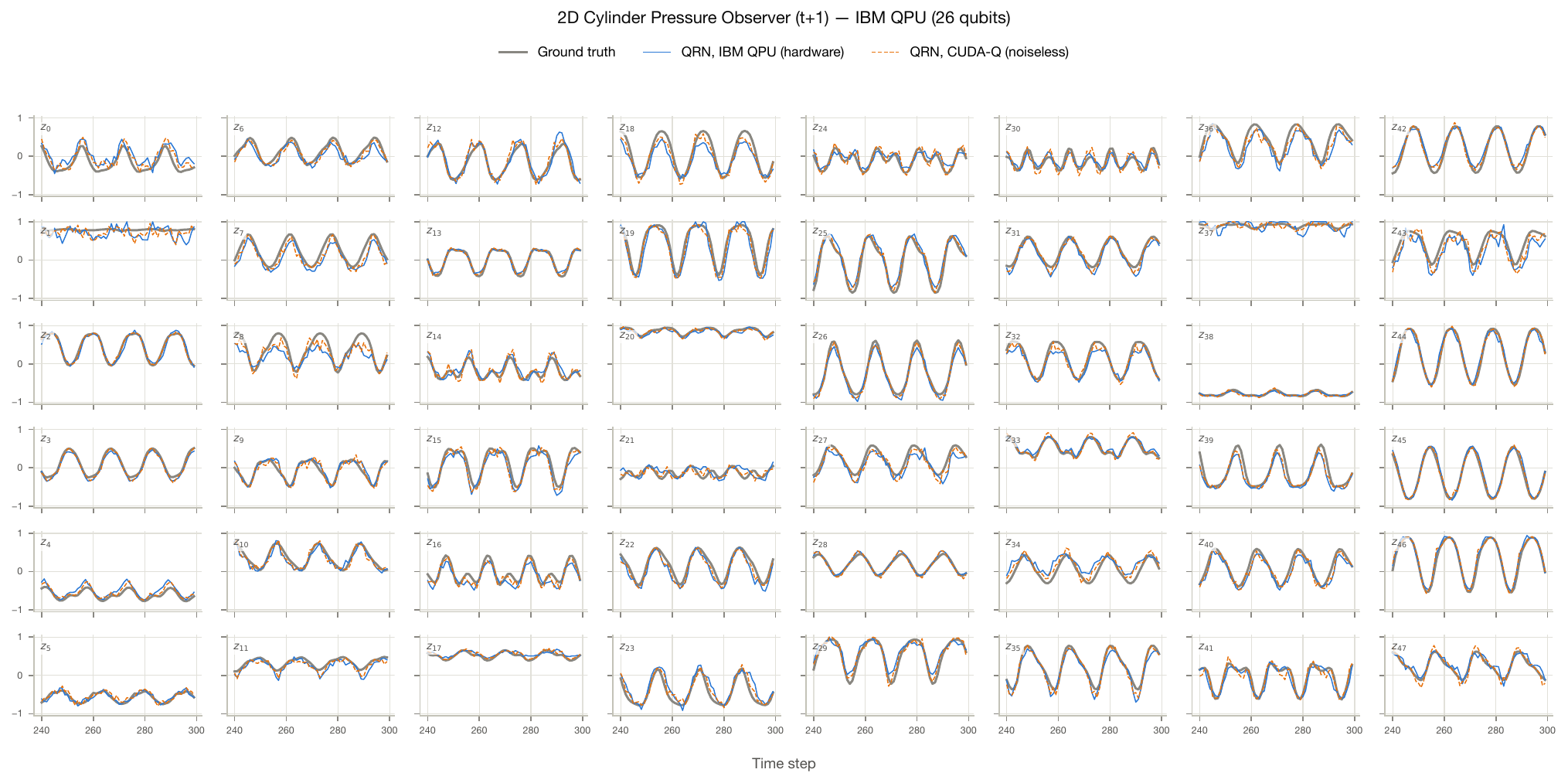}
  \caption{IBM 2D cylinder, pressure-observer ($t{+}1$) task: all 48
  pressure-latent traces over the test window (240--299), predicted
  from the hardware observables and velocity latents alone (blue;
  Supplementary Note~\ref{si:pressure}), with the noiseless CUDA-Q
  prediction overlaid (orange, dashed), against the ground truth
  (gray).}
  \label{fig:pressure-ibm-traces}
\end{figure}

\begin{figure}[!ht]
  \centering
  \includegraphics[width=0.72\textwidth]{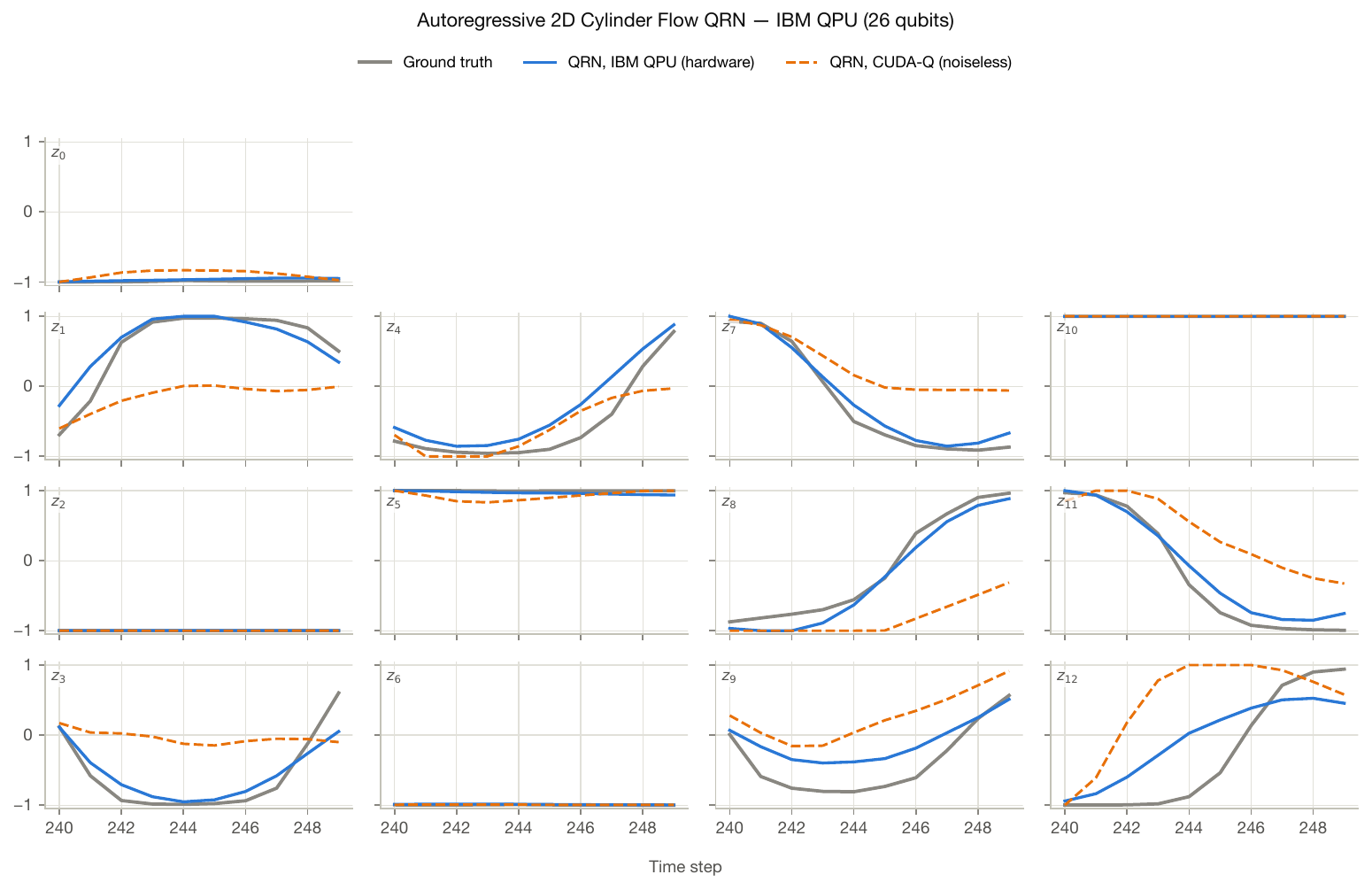}
  \caption{Cylinder latent traces for the 10-step hardware rollout and noiseless CUDA-Q rollout, compared with truth (targets 240--249).}
  \label{fig:ar-cyl-traces}
\end{figure}

\begin{figure}[!ht]
  \centering
  \includegraphics[width=0.72\textwidth]{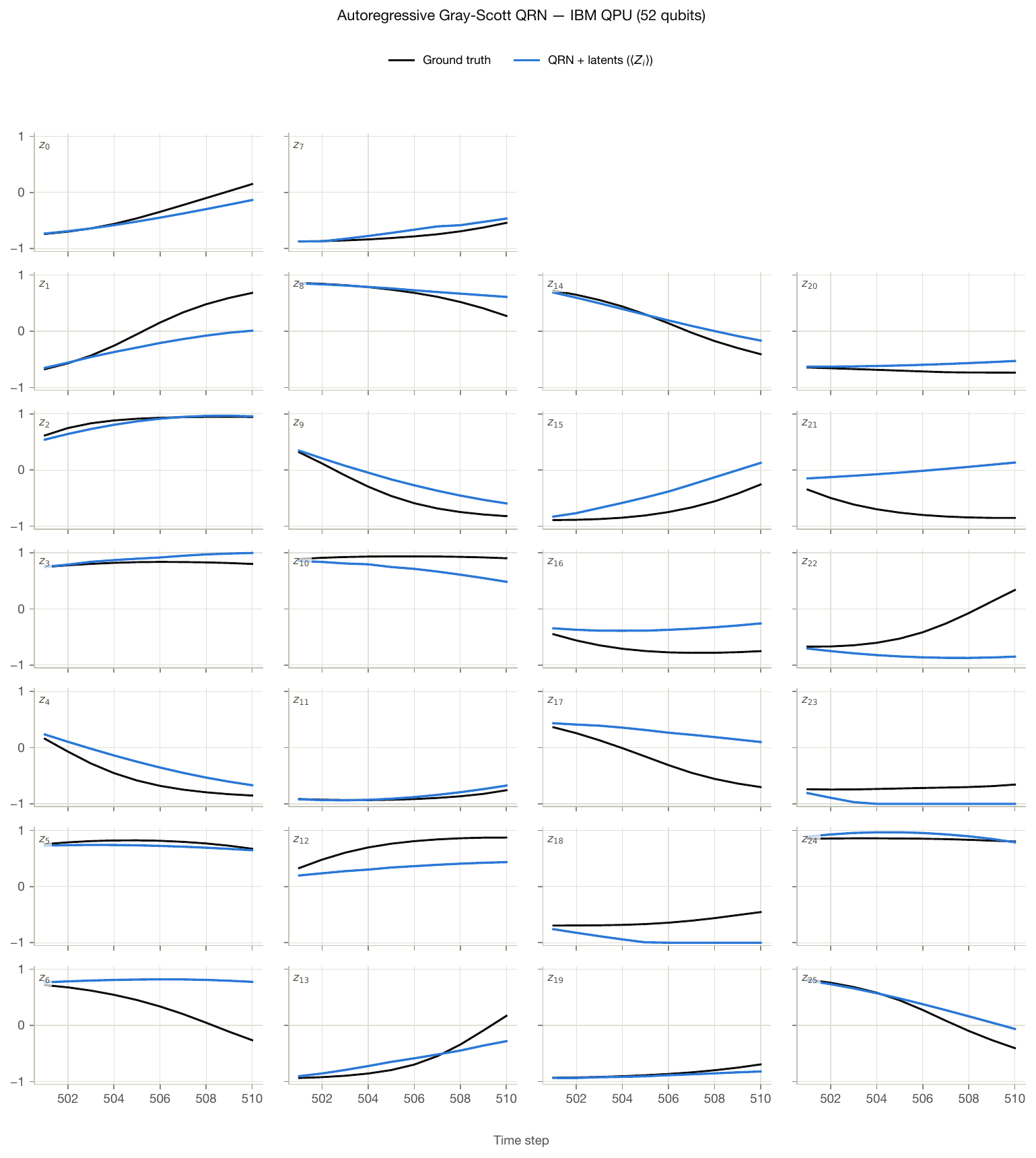}
  \caption{IBM Gray--Scott: per-latent traces of the featured 10-step
  autoregressive hardware rollout (targets 501--510).}
  \label{fig:ar-gs-traces}
\end{figure}

\suppnote{si:traces}{Latent traces}
Latent traces with ground truth and predicted outcomes are detailed in this section in Figs.~\ref{fig:ibm2dcyl-traces}, \ref{fig:gs-traces}, \ref{fig:pressure-ibm-traces}, \ref{fig:ar-cyl-traces}, and \ref{fig:ar-gs-traces}. One-step ($h=1$) prediction traces cover cylinder targets 240--299 and Gray--Scott targets 501--1000. Closed-loop traces cover targets 240--249 and 501--510, respectively. Pressure traces show all 48 observer targets. Dashed CUDA-Q overlays are included where available.

These panels show how aggregate error is distributed across latent
coordinates and time. In one-step tasks, each prediction receives the
true current latent, so an error is not propagated through the input
sequence. The closed-loop traces instead show the combined effects of
prediction and feedback. The close one-step experiment platform agreement can
therefore coexist with visibly different cylinder rollout trajectories.


\begin{figure}[!htbp]\centering
\begin{minipage}[t]{0.49\linewidth}
  \centering
  \includegraphics[width=\linewidth]{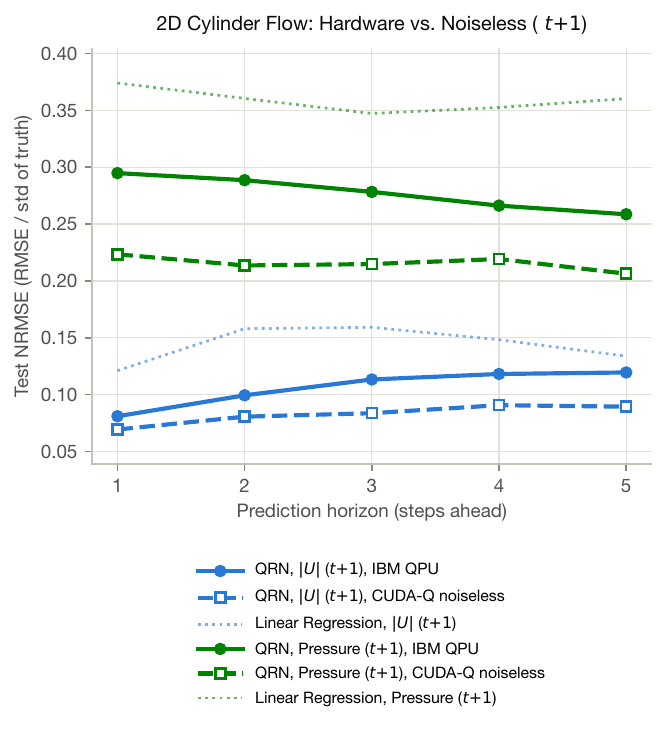}
  \caption{Cylinder direct forecast NRMSE versus horizon for velocity and pressure: IBM and CUDA-Q against each task's linear baseline. Each horizon has its own fitted readout; the supplied plots cover $h=1,\ldots,5$.}
  \label{fig:ibm-vs-cudaq}
\end{minipage}
\hfill
\begin{minipage}[t]{0.49\linewidth}
  \centering
  \includegraphics[width=\linewidth]{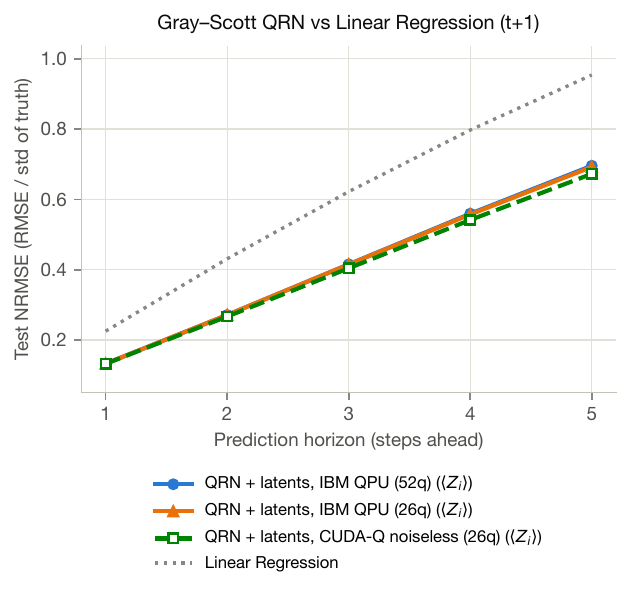}
  \caption{Gray--Scott direct forecast NRMSE versus horizon: 26 qubit CUDA-Q and 26/52 qubit IBM runs against the latent-linear baseline. All use order-1 observables. The 26 qubit hardware reservoir matches the simulation.}
  \label{fig:gs-nrmse}
\end{minipage}
\end{figure}

\suppnote{si:horizons}{Forecast horizons beyond one step}
Here we present results detailing QRN prediction accuracy across different horizons $h$. Fig.~\ref{fig:ibm-vs-cudaq} details IBM vs CUDA-Q performance on the 2D Cylinder tasks, and Fig.~\ref{fig:gs-nrmse} detaile the Gray--scott tasks. A separate readout is used to map the features at $t$ to $z_{t+h}$ for each $h$, without feeding predictions back. Linear baselines use the current latents alone.

The cylinder velocity error increases with horizon, while the pressure
observer is comparatively flat and can improve at later targets. Because
pressure is inferred from velocity rather than extrapolated from its own
past latent, its improved performance over larger horizons may indicate that pressure is a lagging indicator of velocity. For Gray--Scott, the
curves remain close across the three platform/register choices over the
shown horizons, consistent with the small Pauli weight gains in Fig.~\ref{fig:order-ablation}.

\suppnote{si:decoded}{Additional decoded-field comparisons}
The following panels extend the representative reconstructions in the Letter with simulation results and highest-residual hardware frames. Each panel shows truth, prediction, and residual. Field NRMSE includes autoencoder error and is distinct from latent NRMSE.
\begin{figure}[!htb]
  \centering
  \includegraphics[width=\linewidth]{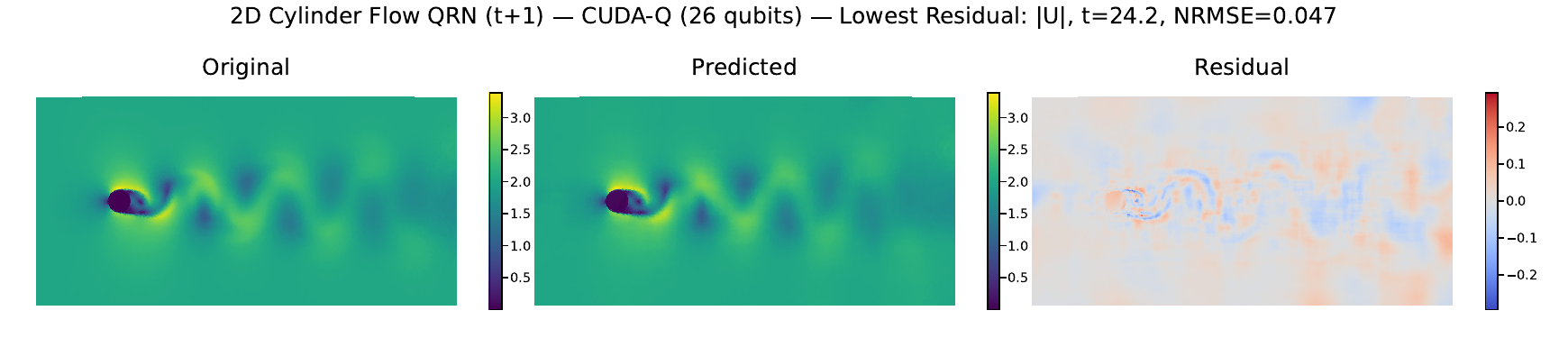}\\[2pt]
  \includegraphics[width=\linewidth]{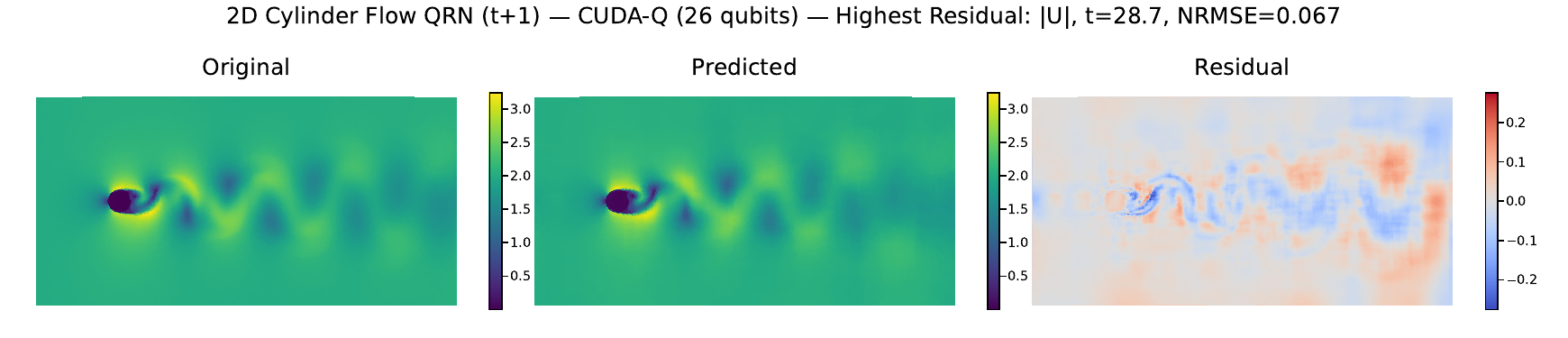}
  \caption{CUDA-Q 2D cylinder, decoded $|U|$ fields (truth /
  prediction / residual): lowest-residual (top) and highest-residual
  (bottom) test frames. The quoted NRMSE is computed against the original (pre-encoding) field and therefore includes the autoencoder reconstruction error.}
  \label{fig:cudaq-decoded}
\end{figure}
\begin{figure}[!htbp]\centering
\includegraphics[width=0.95\linewidth]{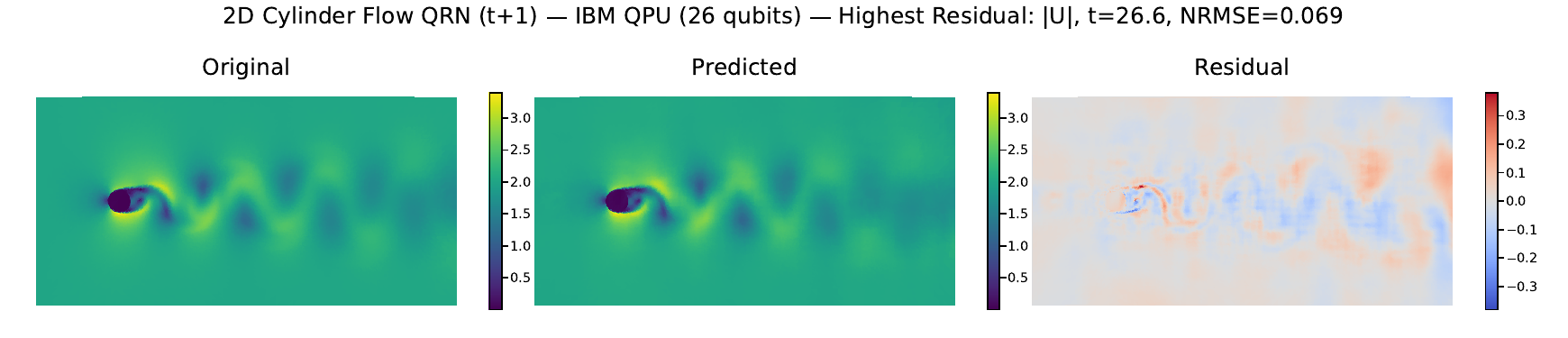}
\caption{Cylinder velocity on IBM: highest-residual one-step test frame.}\label{fig:ibm2dcyl-decoded}
\end{figure}

\begin{figure}[!htbp]\centering
\includegraphics[width=0.95\linewidth]{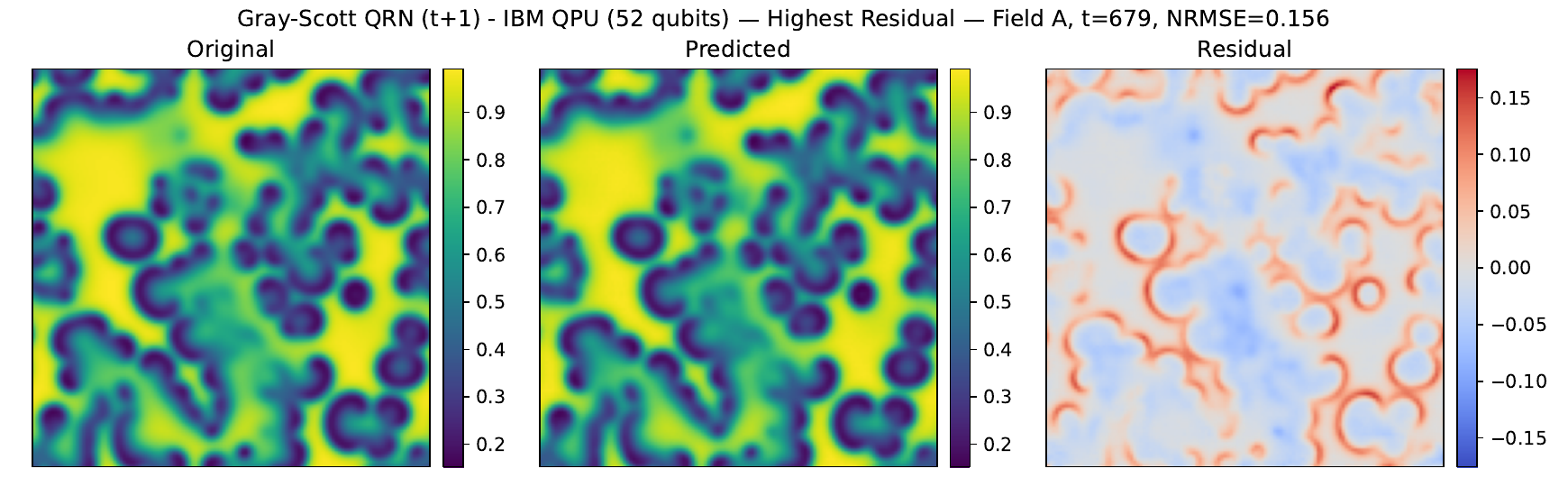}
\caption{Gray--Scott species $A$ on IBM (52 qubits): highest-residual one-step test frame.}\label{fig:gs-decoded}
\end{figure}

\begin{figure}[!htbp]\centering
\includegraphics[width=0.95\linewidth]{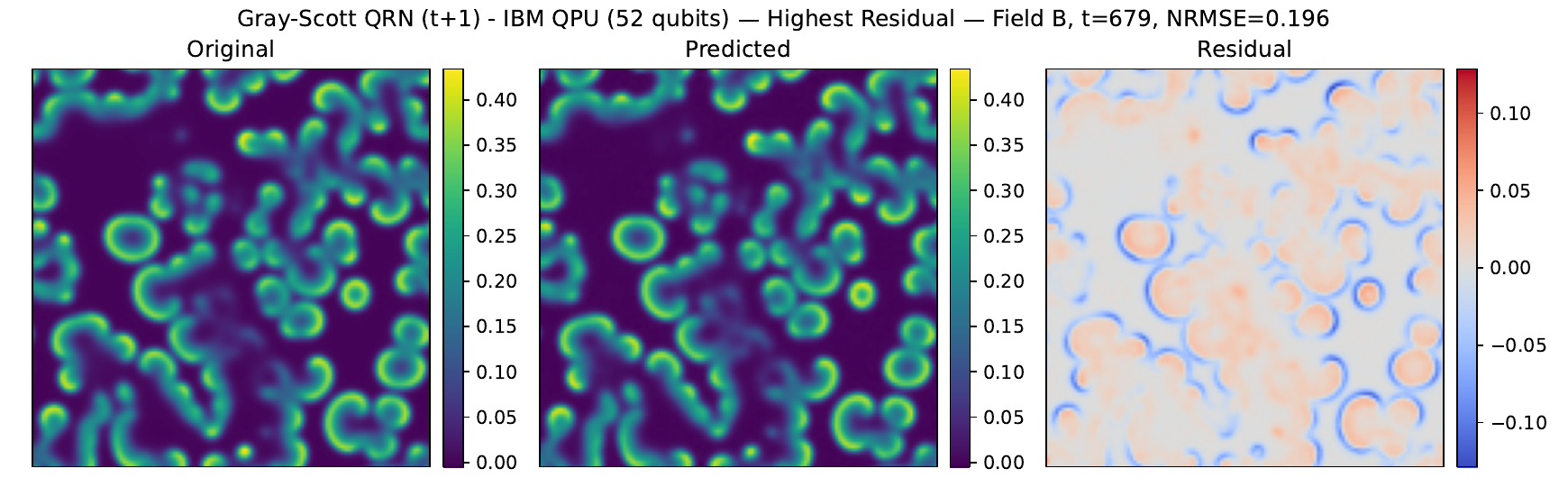}
\caption{Gray--Scott species $B$ on IBM (52 qubits): highest-residual one-step test frame.}\label{fig:gs-decoded-b}
\end{figure}

\FloatBarrier

\end{document}